\documentclass[aps,pre,twocolumn,showpacs,floatfix]{revtex4}
\usepackage{graphicx,graphics,epsf,psfig,epsfig}
\usepackage{subfigure}

\def\beq{\begin{equation}}
\def\eeq{\end{equation}}
\def\beqa{\begin{eqnarray}}
\def\eeqa{\end{eqnarray}}

\begin{document}

\title{Unified perspective on proteins: A physics approach}

\author{Jayanth R. Banavar}
\affiliation{Department of Physics, 104 Davey Lab,
The Pennsylvania State University, University Park PA 16802, USA}

\author{Trinh X. Hoang}
\affiliation{Institute of Physics and Electronics,
VAST, 10 Dao Tan, Hanoi, Vietnam}

\author{Amos Maritan}
\affiliation{INFM and Dipartimento di Fisica `G. Galilei',
Universit\`a di Padova, Via Marzolo 8, 35131 Padova, Italy}

\author{Flavio Seno}
\affiliation{INFM and Dipartimento di Fisica `G. Galilei',
Universit\`a di Padova, Via Marzolo 8, 35131 Padova, Italy}

\author{Antonio Trovato}
\affiliation{INFM and Dipartimento di Fisica `G. Galilei',
Universit\`a di Padova, Via Marzolo 8, 35131 Padova, Italy}

\begin{abstract}
We study a physical system which, while devoid of the
complexity one usually associates with proteins, nevertheless
displays a remarkable array of protein-like properties. The
constructive hypothesis that this striking resemblance is not
accidental leads not only to a unified framework for understanding
protein folding, amyloid formation and protein interactions but
also has implications for natural selection.
\end{abstract}


\maketitle

\newcounter{ctr}
\setcounter{ctr}{1}

\section{Introduction}

The revolution in molecular biology\cite{tbmb} sparked by the
discovery\cite{wc} of the structure of the DNA molecule 50 years
ago has led to a breathtakingly beautiful description of life.
Life employs well-tailored chain molecules to store and replicate
information, to carry out a dizzying array of functionalities and
to provide a molecular basis for natural selection.  The
complementary base pairing mechanism in DNA combined with its
double-helix structure serves as a repository of information and
provides a pretty mechanism for replication\cite{wc}. The
replication is prone to errors or mutations and these errors,
which are the basis of evolution, are in turn copied in future
generations\cite{kimurabook}.  Using the RNA molecule as an
intermediary, the information contained in the DNA genes is
translated into proteins, which are linear chains of amino acids.
Unlike the DNA molecule, which adopts a limited number of related
structures, protein molecules\cite{Fershtbook,Leskbook,Finkelbook}
fold into thousands of native state structures under physiological
conditions.  For proteins, form determines functionality and the
rich variety of observed forms underscores the versatility of
proteins.  There then follows a complex orchestrated dance in
which proteins catalyze reactions, interact with each other and
finally feedback into the gene to regulate the synthesis of other
proteins\cite{tbmb}.

A protein molecule is large and has many atoms.  In addition, the
water molecules surrounding the protein play a crucial role in
its behavior.  At the microscopic level, the laws of quantum
mechanics can be used to deduce the interactions but the number
of degrees of freedom are far too many for the system to be
studied in all its detail.  When one attempts to look at the
problem in a coarse-grained manner\cite{bmedit} with what one hopes are the
essential degrees of freedom,  it is very hard to determine what
the effective potential energies of interaction are.  This
situation makes the protein problem particularly daunting and
no solution has yet been found.

Over many decades, much experimental data has been accumulated yet
theoretical progress has been somewhat limited.  The problem is highly
interdisciplinary and touches on biology, chemistry and physics and it
is often hard to distill the essential features of each of the
multiple aspects of the problem. The great successes of quantum
chemistry in the determination of the structure of the DNA
molecule\cite{wc} and in the spectacular prediction that helices and
sheets\cite{Pauling1,Pauling2,EisenbergPNAS} are the building blocks
of protein structures has spurred much work using detailed chemistry
on understanding the protein problem.  Such work has been very
insightful in providing useful hints on how proteins behave at the
atomic scale in performing their tasks.  The missing feature, of
course, in such a theoretical approach is that it treats each protein
as a special entity with all the attendant details of the sequence of
amino acids, their intricate side chain atoms and the water molecules.
Such an approach, while quite valuable, neither has as a goal nor can
lend itself to a unified way of understanding seemingly disparate
phenomena pertaining to proteins.  Reinforcing this, experiments,
which are very challenging, are carried out on one protein at a time
and cry out for an understanding of the behavior of an individual
class of protein.

The lessons we have learned from physics are of a different
nature. The history of physics is replete with examples of the
elucidation of connections between what seem to be distinct
phenomena and the development of a unifying framework, which, in
turn, leads to new observable
consequences\cite{history,Dysonbook}. There have been many
attempts at using physics-based approaches for understanding
proteins. These have provided valuable insights on how one might
think about the problem and have served as a means of
understanding experimental data. Yet, no simple unification has
been achieved in a deeper understanding of the key principles at
work in proteins.

We restrict ourselves to globular proteins which display the rich
variety of native state structures. There are other interesting
and important classes of proteins\cite{Creighton} such as membrane
proteins and fibrous proteins which we do not consider here.
Our goal here is to present a new approach to understanding
proteins -- our focus is on understanding the origin of protein
structures and how they form the basis for both functionality and
natural selection. Our work points to a unification of the various
aspects of all proteins: symmetry and geometry determine the
limited menu of folded conformations that a protein can choose
from for its native state structure; these structures are in a
marginally compact phase in the vicinity of a phase transition and
are therefore eminently suited for biological function; these
structures are the molecular target for the powerful forces of
evolution; proteins are well-designed sequences of amino acids
which fit well into one of these predetermined folds; and proteins
are prone to misfolding and aggregation leading to the formation
of amyloids, which are implicated in debilitating human
diseases\cite{Kelly,Ddisease} such as Alzheimer's, light-chain
amyloidosis and spongiform encephalopathies.

We present a discussion of the nature of the denatured state (which
can loosely be thought of as the collection of unfolded conformations)
and its possible key role in the protein folding problem.  We also
show how disordered proteins could fit into our unified framework.

The problem of how life was created is a fascinating one. Our
focus is on looking at life on earth and asking how it works.  The
lessons we learn provide hints to the answers of deep and
fundamental questions that have been pondered by our ancients: Was
life on earth inevitable? Then there is the question posed by
Henderson\cite{Henderson} about whether the nature of our physical
world is biocentric? Is there a need for fine-tuning in
biochemistry to provide for the fitness of life in the cosmos or
even less ambitiously for life here on earth? Surprisingly, as we
will show, a physics approach turns out to be valuable for
thinking about these questions.

The main text of the paper contains the principal ideas and
details of the calculations are relegated to the appendices. In
section II, we introduce the description of a protein as a thick
polymer chain and highlight the differences in its phase diagram
with respect to the usual string and beads model. In section III,
we make a comparison of the predictions obtained from the simple
tube model against experimental data available on protein native
state structures. In section IV, we introduce a more refined model
in which the tube picture is reinforced with the geometrical
constraints that arise in the formation of hydrogen-bonds and
discuss the resulting phase diagram for an isolated peptide chain.
In section V, we discuss several consequences of our model
including the nature of the free energy landscape, the innate
propensity of proteins to aggregate into amyloid-like forms and
the role played by proteins as the targets of natural selection in
molecular evolution. In section VI, we discuss the nature of the
denatured state of proteins and its possible role in protein
folding.  In the final section VII, we conclude with a summary.

\section{Phases of matter: from spheres to tubes}

The fluid and crystalline phases of matter can be readily
understood\cite{hardsphere} in terms of the behavior of a simple
system of hard spheres. The standard way of ensuring the
self-avoidance of a system of uniform hard spheres is to consider all
pairs of spheres and require that their centers are no closer than
their diameter. Studies of hard spheres have a venerable
history\cite{Szpiro} including early work by Kepler on the packing of
cannonballs in a ship's hold.  Each hard sphere can be thought of as a
point particle or a zero dimensional object with its own private space
of spatial extent equal to its radius.  Generalizing to a one
dimensional object, one must consider a line or a string, with private
space associated with each point along the line, leading to a uniform
tube of radius of cross-section or thickness, $\Delta$, with its axis
defined by the line. (Likewise, one could consider a collection of
interacting tubes.)  The generalization of the hard sphere constraint
to the description of the self-avoidance of a tube of non-zero
thickness is as follows\cite{JSP} (see Appendix A): consider all
triplets of points along the axis of the tube.  Draw circles through
each of the triplets and ensure that none of the radii is less than
the tube thickness\cite{GM}.  This prescription surprisingly entails
discarding pairwise interactions and working with effective three body
interactions\cite{JSP,CPU,Macro}.

\begin{figure}
\centering
\epsfig{width=8.0cm,file=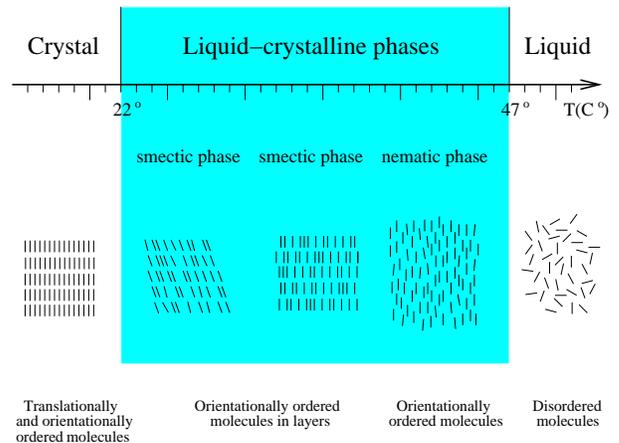}
\caption{(Color online) Schematic phase diagram for hard rods highlighting the rich
behaviour and the new (with respect to hard spheres) liquid crystal
phases exhibited at intermediate temperatures.}
\label{Fig1}
\end{figure}

One may visualize a tube as the continuum limit of a {\em
discrete} chain of tethered disks or coins\cite{CPU} of fixed
radius separated from each other by a distance $a$ in the limit of
$a \rightarrow 0$. The inherent anisotropy associated with a coin
(the heads to tails direction being different from the other two
perpendicular to it) reflects the fact that there is a special
local direction at each position defined by the locations of the
adjacent objects along the chain.  An alternative description of a
discrete chain molecule is a string and beads model in which the
tethered objects are spheres.  The key difference between these
two descriptions is the different symmetry of the tethered
objects.  Upon compaction of a chain of spheres, each
individual sphere tends to surround itself isotropically with
other spheres unlike the tube situation in which nearby tube
segments need to be placed parallel to each other. Even for
unconstrained particles, deviations from spherical symmetry
(replacing a system of hard spheres with one of hard rods, for
example) lead to rich new liquid crystal
phases\cite{DeGennes,Chandra} (see Fig. \ref{Fig1}).  Likewise, we
find that the tube and a chain of tethered spheres exhibit quite
distinct behaviors with one exception -- in the presence of an
attractive self-interaction favoring compaction, the chain of
coins and the string and beads model behave similarly in the limit
of vanishing ratios of the radii of the coin and sphere to the
range of attraction. A detailed comparison between the chain of
coins (tube) and the string and beads model with a bending
rigidity energy term is carried out in Appendix B.

\begin{figure}
\centering \epsfig{width=8.0cm,file=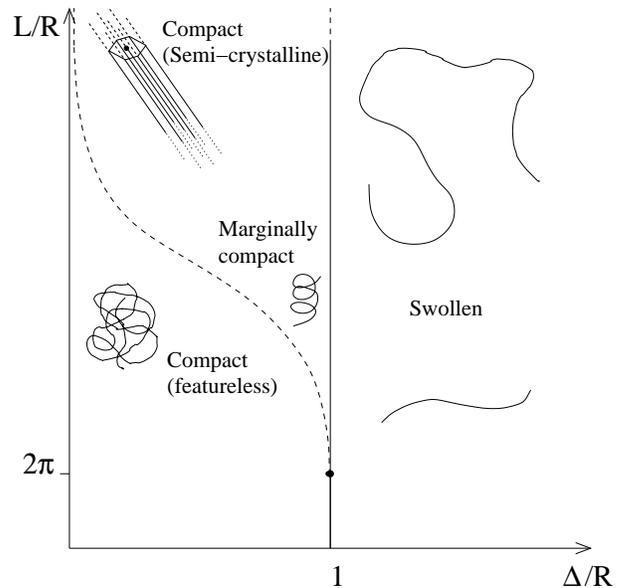} \caption{Sketch of
the zero temperature phase diagram of a tube in the continuum,
subject to a self-attraction promoting compaction. There are two
phases when the tube length $L$ is long compared to the range of
attractive interaction $R$. One obtains the semi-crystalline phase
(with parallel/anti-parallel alignment between different stretches
of the tube which then fill the space with hexagonal symmetry, as
depicted in the figure) when the tube thickness $\Delta$ is small
compared to $R$ and a swollen phase when $\Delta$ is large
compared to $R$.  There are interesting finite size effects in the
semi-crystalline phase. In the thin tube limit, on decreasing the
length there is a crossover from the semi-crystalline phase with
overall cylindrical symmetry to a featureless compact phase with
spherical symmetry when $L/R \sim (\Delta/R)^{-2}$. There is an
unusual finite size effect when $\Delta \sim R$ near the
confluence of three phases (at $L=2\pi R$, $\Delta=R$ for a chain
in the continuum): the semi-crystalline phase, the featureless
compact phase and the swollen phase. A marginally compact phase is
obtained in this regime and displays a dramatic entropy reduction,
with the choice structure being a helix with a well defined pitch
to radius ratio (see Fig. \ref{Fig4}). Other structures such as
hairpins and sheets are present in the marginally compact phase
for {\em discrete chains} (see Fig. \ref{Fig3})\cite{Macro}.}
\label{Fig2}
\end{figure}

Fig. \ref{Fig2} is a sketch of the phase diagram, at zero
temperature, of a homopolymer of length $L$ and thickness $\Delta$
with the range of attractive interaction $R$.  This phase diagram
has been obtained using detailed computer simulations accompanied
by an approximate mean field theory\cite{Macro} and can be
understood on the basis of physical arguments.  For large values
of $L/R$, there are two distinct phases. When $\Delta/R$ is large,
the tube is very thick compared to the range of attractive
interactions and one obtains a swollen phase with equal weight for
all self-avoiding conformations.  One finds a very large
degeneracy with no tendency towards compaction.  On the
other hand, for small $\Delta/R$, one has a semi-crystalline
phase\cite{semicrystal} in which the tube is stretched out locally
with nearby sections parallel to each other.  A similar structure
is also obtained for many long tubes -- the arrangement is akin to
piling up logs parallel to each other with each log surrounded by
six other logs in a hexagonal array, the optimal packing in two
dimension of coins of radius $\Delta$.  Such structures are
similar to those found in the Abrikosov flux
lattice\cite{Abrikosov} and bear a resemblance to liquid crystal
order.

Liquid crystals are a delicate state of matter of rod-like molecules
which adopt many distinct arrangements sensitive to external electric
and magnetic fields\cite{DeGennes,Chandra}.  A liquid crystal phase
that is analogous to the semi-crystalline phase is the nematic phase
in which the molecules move as in a regular liquid but with an
alignment of their axes.  Unlike a spin system in which an up spin is
different from a down spin, in the nematic phase, all that matters is
the direction of the axis of the particle -- there is no up-down
distinction -- and this change in symmetry leads to a first order
phase transition between the disordered isotropic and the ordered
nematic phases. Likewise, the phase transition between the
semi-crystalline phase at low temperatures and a high temperature
disordered phase in which there is no compaction of the tube is a
first order transition as in the melting of ice into water.
At the transition temperature, there is a coexistence of the two
phases (e.g., pieces of ice floating in a glass of water) and an
abrupt transition between the two states.  One might call such a
system a two-state system -- one has water and/or ice but nothing in
between.

When the tube is short, one would expect finite size
effects\cite{Clusters} to come into play.  In most physical systems,
such finite size effects are intuitively obvious corrections to the
bulk scenario and arise from the effects of the finite boundaries.
For our tube, the simplest situation occurs in the swollen phase where
the finite size effects are not important -- short fat tubes continue
to adopt open conformations.  At the other extreme of small
$\Delta/R$, as one reduces the length of the tube, the overall
symmetry of the folded object crosses over from that of a cylinder
(corresponding to the Abrikosov flux lattice-like phase akin to the
hexagonal arrangement of parallel, straight logs) to a sphere when $L
\sim R^3/\Delta^2$ and one obtains one out of many degenerate
featureless compact conformations.  Physically, for a short tube,
there are many more conformations that can be accommodated in the
spherical topology than in the cylindrical topology without any
accompanying sacrifice in the attractive interaction energy.

\begin{figure}
\centering \epsfig{width=3.2in,file=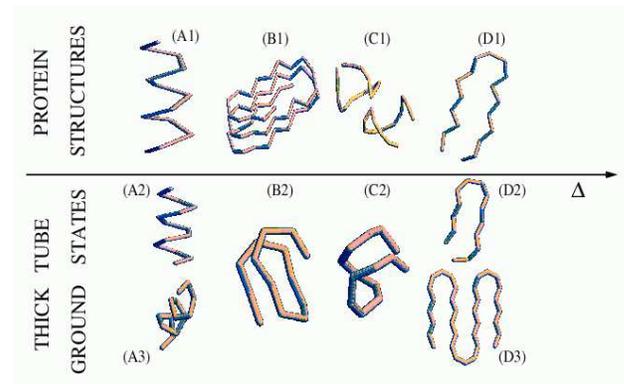}
\caption{(Color online) Building blocks of biomolecules and ground state structures
associated with the marginally compact phase of a short tube
corresponding to a discrete chain of tethered disks of radius
$\Delta$. The axis in the middle indicates the direction along which
the tube thickness $\Delta$ increases.  The top row shows some of the
building blocks of biomolecules, while the bottom row depicts the
corresponding structures obtained as the ground state conformations of
a short tube.  (A1) is an $\alpha$-helix of a naturally occurring
protein, while (A2) and (A3) are the helices obtained in our
calculations -- (A2) has a regular contact map (i.e. a matrix whose
elements, corresponding to residue pairs, are either $0$ or $1$
depending on whether the two given residues are in contact or not)
whereas (A3) is a distorted helix in which the distance between
successive atoms along the helical axis is not constant but has period
$2$. (B1) is a helix of strands in the alkaline protease of {\em
pseudomonas aeruginosa}, whereas (B2) shows the corresponding
structure obtained in our computer simulations. (C1) shows the
``kissing'' hairpins of RNA and (C2) the corresponding conformation
obtained in our simulations.  Finally (D1) and (D2) are two instances
of quasi-planar hairpins.  The first structure is from the same
protein as before (the alkaline protease of {\em pseudomonas
aeruginosa}) while the second is a typical conformation found in our
simulations. The sheet-like structure (D3) is obtained for a longer
tube (see \cite{CPU} for more details). The biomolecular structures
in the top row are shown in the $C^{\alpha}$ representation for
proteins, and in the $P$ representation for RNA kissing hairpins.}
\label{Fig3}
\end{figure}

\begin{figure}
\centering \hfill
\subfigure[]{\includegraphics[height=4.1cm]{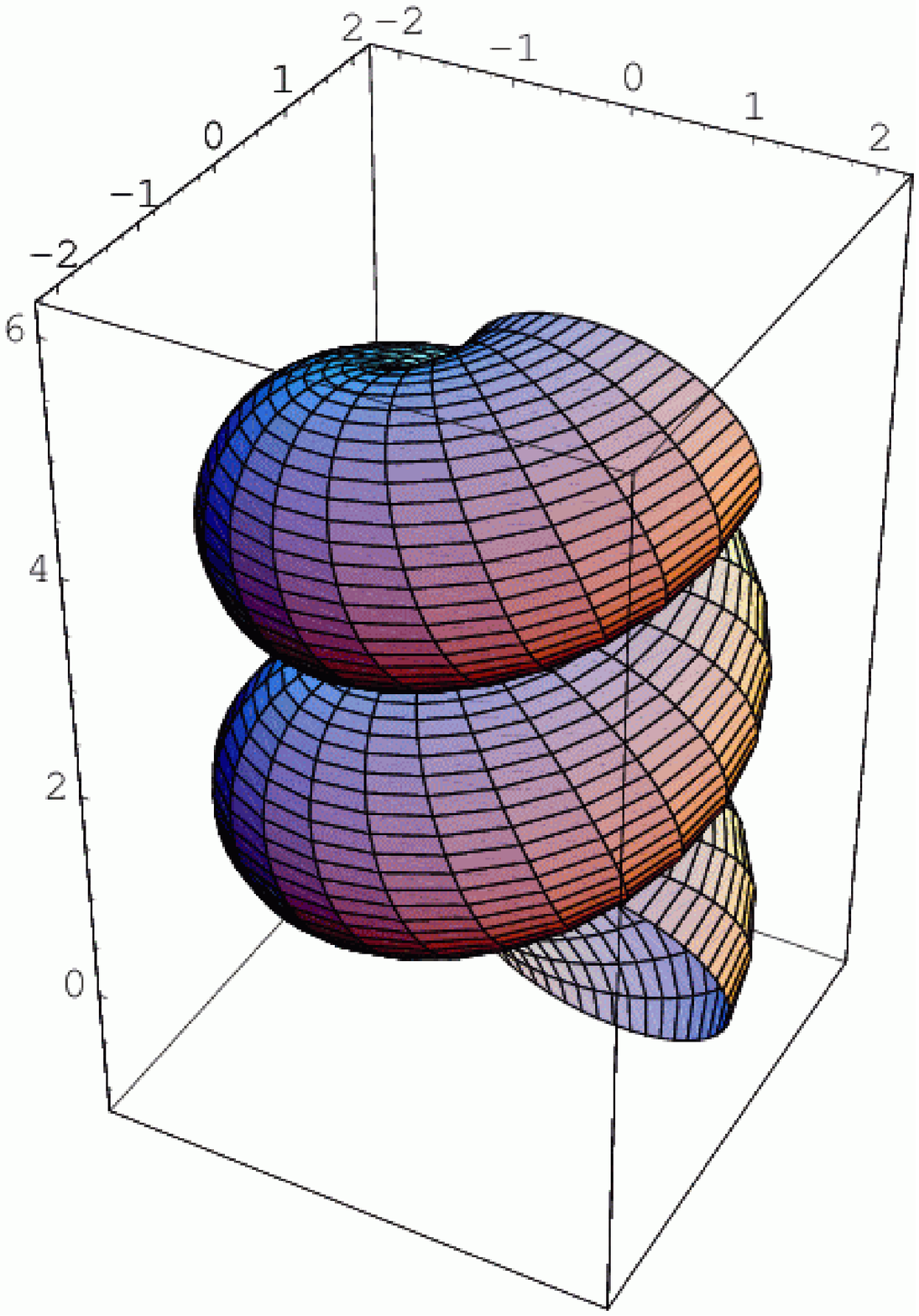}} \hfill
\subfigure[]{\includegraphics[height=4.7cm]{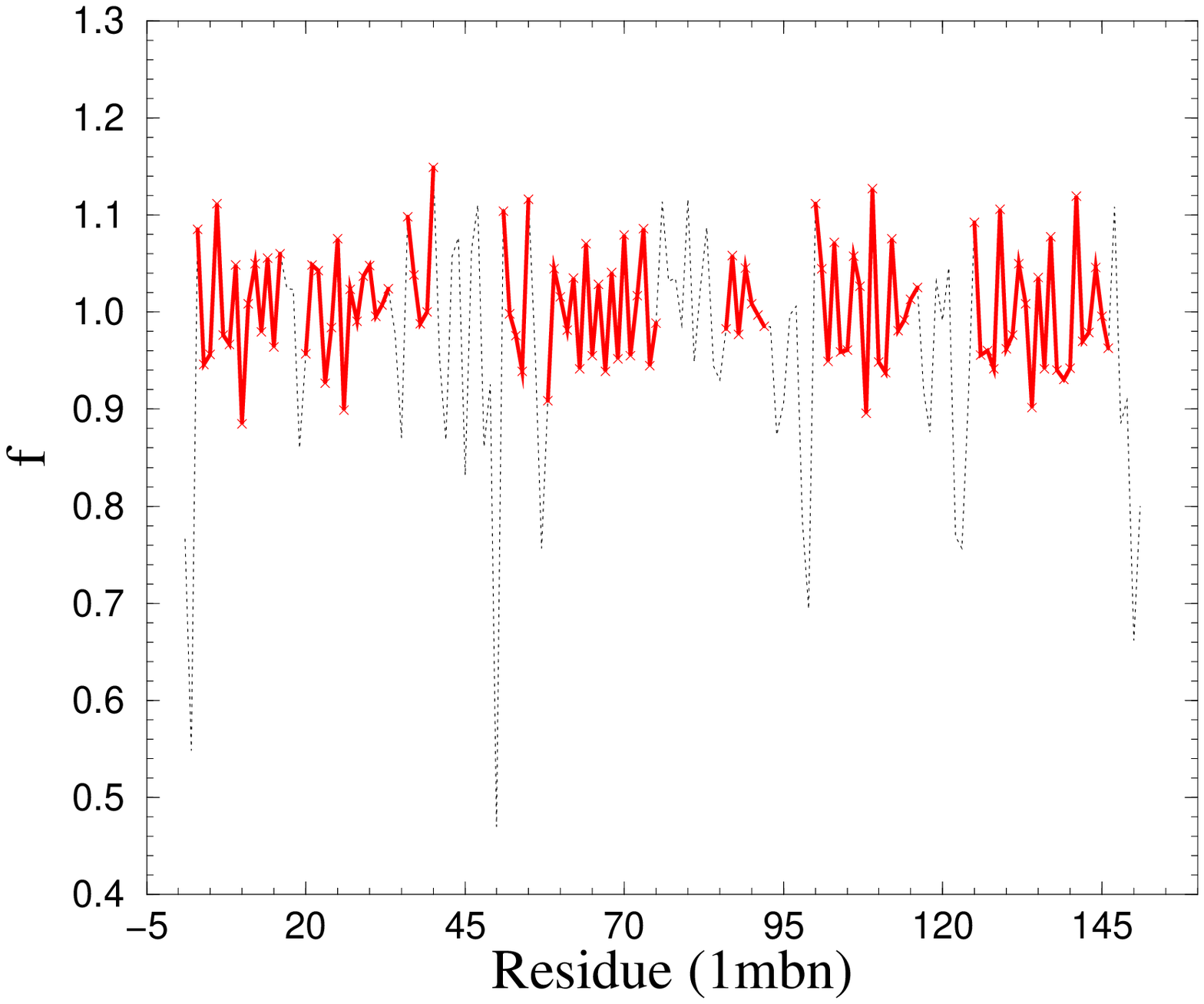}} \hfill
\caption{(Color online) (a) Space filling {\em optimal} helix, with a pitch to radius
ratio $c^* \approx 2.512$ (drawn using Mathematica). As explained in
Appendix C, this optimal value is determined by requiring that the
radius of curvature of the helical curve is equal to half the minimum
distance of closest approach between different turns of the helix. The
corresponding tube (that can be thought of as being inflated uniformly
around the curve) is optimally space filling since it stops growing
when reaching its maximum thickness both {\em locally} (the radius of
curvature) and {\em non locally} (half the minimum distance of closest
approach between different turns) at the same time (see Appendices A
and C). Such an optimality criterion is shared by some of the
conformations selected as ground states in our simulations in the
marginally compact phase such as helices or planar hairpin and sheets
shown in Fig. \ref{Fig3}, when it is properly translated for the case
of a discrete chain (see below and eq. (\ref{Disc}) in Appendix A). It
can be shown\cite{CPU} that the planarity of hairpins and sheets is a
consequence of this optimal space-filling criterion. The same
geometrical feature is strikingly found to hold, within $3\%$, for
$\alpha$-helices occurring in native state of natural
proteins\cite{MaritanNature}. (b) Plot of the ratio
$f_i=\rho_{NL}(i)/\rho_L(i)$ of the non-local radius of curvature
$\rho_{NL}(i)\equiv\min_{j<k}r({\bf r}_i,{\bf r}_j,{\bf r}_k)$ (with
$\left\{j,k\right\}\ne\left\{i-2,i-1\right\},
\left\{i-1,i+1\right\},\left\{i+1,i+2\right\}$) over the radius of
curvature $\rho_L(i)\equiv r({\bf r}_{i-1},{\bf r}_i,{\bf r}_{i+1})$
as a function of the residue index $i$ for the native state structure
of sperm whale myoglobin (PDB code 1mbn), where ${\bf r}_i$ refers to
the spatial coordinates of the $C^{\alpha}$ atom of the $i$-th
residue, $1\le i\le153$ (see Appendix A for the definition of the
triplet radius $r({\bf r}_i,{\bf r}_j,{\bf r}_k)$). In correspondence
with the $8$ $\alpha$-helices present in the myoglobin fold, shown as
the solid (red) parts in the plot, the values of $f_i$ oscillate
around unity, demonstrating that helices in natural proteins are {\em
optimally} space filling in the sense described above.}
 \label{Fig4}
\end{figure}

There is a confluence of three distinct types of structures: the
swollen conformations, the semi-crystalline phase and the
featureless compact conformations, when $\Delta \sim R \sim L$
(Fig. \ref{Fig2}). This interplay leads to quite remarkable finite
size effects: one obtains a {\em marginally compact} phase with a
huge reduction in the degeneracy compared to the featureless
compact phase and the swollen phase.  On raising the temperature,
one again finds a two-state behavior and the finite size analog of
a first order transition between the marginally compact phase and
the disordered phase.  The first order transition occurs because
it is necessary for different nearby tube segments to snap into
position right alongside each other and parallel to each other in
order to avail of the attraction.  The inherent anisotropy of a
tube along with the fact that $\Delta$ is of order $R$ leads to
this requirement.  Such two-state behavior can, in the simplest
scenario, be associated with a transition state\cite{Eyring} along
suitably chosen reaction coordinates.  The structures of
choice\cite{CPU,BanavarRMP} in the marginally compact phase, for a
{\em discrete chain}, are helices, kissing hairpins, regular
hairpins and sheets (Figures \ref{Fig3} and \ref{Fig4}). Helices,
hairpins and sheets are indeed characterized by a parallel
placement of nearby tube segments.  The marginally compact phase
is poised in the vicinity of a phase transition to the swollen
phase and the structures are therefore flexible\cite{Thorpe123}
and sensitive to the right types of perturbations.

\section{Tubes and proteins}

There is a truly remarkable coincidence between the structures one
obtains in the marginally compact physical state of matter of
short tubes and the building blocks of protein native state
structures (Fig. \ref{Fig3}).  Proteins\cite{Creighton} are linear
chains of amino acids, of which there are twenty naturally
occurring types with distinct side chains. The backbone and
several of the side chains are hydrophobic and, under
physiological conditions, globular proteins fold rapidly and
reproducibly to somewhat compact conformations called their native
state structures. In their native states, a hydrophobic core is
created which is space-filling and water is expelled from the
interior.  Even though there are hundreds of thousands of proteins
in human cells, the total number of distinct folds that they adopt
in their native states is only of the order of a few
thousand\cite{ChFin90,Chothia1,PrytzkaRose}.  Furthermore, these
structures seem to be evolutionarily
conserved\cite{Chothia2,Denton,Koonin}. Proteins are relatively
short chain molecules and indeed longer globular proteins form
domains which fold autonomously\cite{MD}.  The building blocks of
protein structures are helices, hairpins and almost planar sheets
(Fig. \ref{Fig3}). Strikingly, short tubes, with no {\em
heterogeneity}, in the marginally compact phase form helices with
the same pitch to radius ratio as in real
proteins\cite{MaritanNature} (Fig. \ref{Fig4}) and almost planar
sheets made up of zig-zag strands. It is interesting to note that
the helix is a very natural conformation for a tube and occurs
without any explicit introduction of hydrogen bonding.  Recent
work on the denatured state of short amino acid sequences has
suggested that the poly-proline II helix might be the preferred
structure in that phase, even though it does not entail the
formation of any hydrogen bonds\cite{polyproline}. As in the tube
case, small globular proteins show a two-state
behavior\cite{Ginsburg,Anfinsen,Jackson,BaldwinRose,Baker} and
recent experiments\cite{Fershtbook,TS1,TS2} have been successful
in mapping out the nature of the transition state in several
cases.

Let us make the {\em constructive hypothesis} that the extraordinary
similarity between the structures adopted by short tubes in the
marginally compact phase and the building blocks of protein native
state structures is not a mere coincidence. We {\em postulate} instead
that the tube picture presented above is a paradigm for understanding
protein structures.  Quite generally, such postulates are of limited
utility unless one is able to unify seemingly unrelated aspects of the
problem and make new predictions amenable to experimental
verification.  In our case, while the tube idea is theoretical, there
is a wealth of experimental data already available on proteins.
Before we proceed to explore the consequences of our hypothesis, we
will first link the tube picture with the protein problem using
experiments as the guide.

\begin{figure}
\centering \epsfig{height=8.0cm,,angle=-90,file=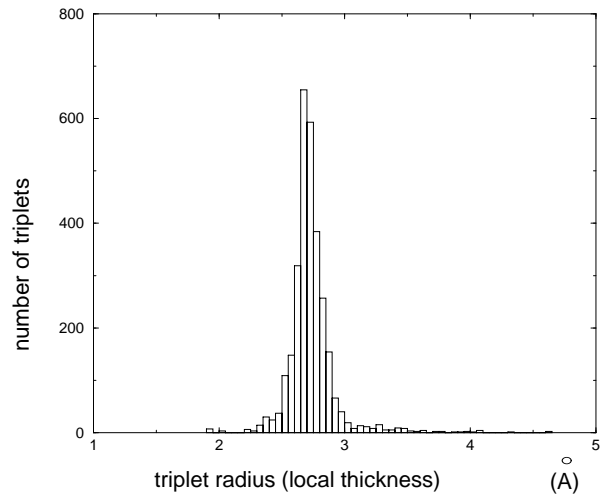}
\caption{Histogram of local thicknesses computed for all residues
of different protein native structures, when the virtual chain
formed by the backbone $C^{\alpha}$ atoms is viewed as a
discretized thick tube. At a given residue the local thickness is
simply the minimum triplet radius over all triplets containing
that residue (see Appendix A for the definition of triplet radius
and for an explanation of how such a quantity arises within a tube
description).} \label{Fig5}
\end{figure}

Let us begin by asking whether the backbone of a protein can be
described as a tube.  Fig. \ref{Fig5} indeed shows that, in its native
state, the protein backbone can be thought of as the axis of a tube of
approximate radius of cross-section ($\Delta$) equal to
$2.7$ \AA. Interestingly, there are small variations in the tube radius
especially in the vicinity of backward bends\cite{BMMTO2}.  The tuning
of the two length scales, $\Delta$ and $R$, to be comparable to each
other happens automatically for proteins: the sizes of the amino acid
side chains determine both the tube thickness and the range of
interactions.  Steric interactions lead to a vast thinning of the
phase space that protein structures can explore\cite{Rama,Linus}.
Physically, the notion of a thick chain or a tube follows directly
from steric interactions in a protein -- one needs room around the
backbone to house the amino acid side chains without any overlap.  The
same side chains that determine the tube thickness also control the
range of attraction -- the outer atoms of the side chain interact
through a short-range interaction screened by the water.  This
self-tuning is a quite remarkable feature of proteins.

The rapid folding of small proteins can be understood in terms of the inherent
anisotropy of a tube and the self-tuning of the two key length
scales, the tube thickness and the range of the attractive interactions.
In the marginally compact phase, in order to avail of the attractive
interactions, nearby segments of the tube have to snap into place parallel
to each other and right up against each other.  As stated before,
both in the tube picture and in proteins,   the helix and
the sheet are characterized by such parallel space-filling alignment of
nearby tube segments.  In proteins, such an arrangement serves to
expel the water from the protein core. As shown by Linus
Pauling and coworkers\cite{Pauling1,Pauling2},
hydrogen bonds provide the scaffolding
for both helices and sheets and place strong geometrical constraints
stemming from quantum chemistry.

\section{Beyond the tube archetype: A refined tube model informed by protein data}

\begin{figure}
\centering
\epsfig{height=8.0cm,file=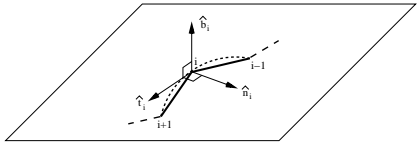,angle=-90}
\caption{Sketch of the local coordinate system. For each $C^{\alpha}$
atom $i$ (except the first and the last one), the axes of a
right-handed local coordinate system are defined as follows. The
tangent vector $\hat t_i$ is parallel to the segment joining $i-1$
with $i+1$. The normal vector $\hat n_i$ joins $i$ to the center of
the circle passing through $i-1$, $i$, and $i+1$ and it is
perpendicular to $\hat t_i$. $\hat t_i$ and $\hat n_i$ along with the
three contiguous $C^{\alpha}$ atoms lie in a plane shown in the
figure. The binormal vector $\hat b_i$ is perpendicular to this
plane. The vectors $\hat t_i$, $\hat n_i$, $\hat b_i$ are normalized
to unit length.}
\label{Fig6}
\end{figure}

\begin{figure}
\centering \epsfig{width=3.2in,file=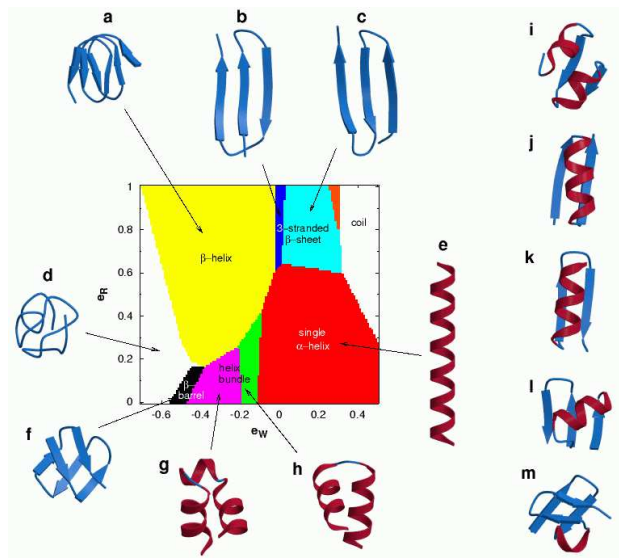} \caption{(Color online) Phase
diagram of ground state conformations.  The ground state
conformations were obtained by means of Monte-Carlo simulations of
chains of 24 $C^{\alpha}$ atoms. $e_R$ and $e_W$ denote the local
radius of curvature energy penalty and the solvent mediated
interaction energy respectively (see Appendix E). Over $600$
distinct local minima were obtained in our simulations in
different parts of parameter space starting from a randomly
generated initial conformation. The temperature is set initially
at a high value and then decreased gradually to zero.
(a), (b), (c), (e), (f), (g), (h) are the Molscript representation of
the ground state conformations which are found in different parts of
the parameter space as indicated by the arrows. The helices and
strands are assigned when local or non-local hydrogen bonds are formed
according to the rules described in Appendix E. Conformations (i),
(j), (k), (l), (m) are competitive local minima. In the orange phase,
the ground state is a 2-stranded $\beta$-hairpin (not shown). Two
distinct topologies of a 3-stranded $\beta$-sheet (dark and light blue
phases) are found corresponding to conformations shown in
conformations (b) and (c) respectively.
The white region in the left of the phase diagram has large attractive
values of $e_W$ and the ground state conformations are compact
globular structures with a crystalline order induced by hard sphere
packing considerations\cite{Karplus} and not by hydrogen bonding
(conformation (d)).}
\label{Fig7}
\end{figure}

We turn now to a marriage of the tube idea and the wealth of
information available from a variety of experimental
probes\cite{Fershtbook,EXPTS} in preparation for the task of exploring
the consequences of our hypothesis.  Recall that three body local and
non-local radii constraints describe the self-avoidance of a
tube\cite{JSP} (see Appendix A). For a discrete chain, the local three
body radius is defined as the radius of a circle drawn through three
consecutive nodes of the chain (in the limit of a continuous chain the
local three body radius is equal to the radius of curvature). The
non-local radius at a given node is defined to be the smallest among
all the radii of circles drawn through that node and all pairs of
other nodes except for its adjacent nodes (see also the caption of
Fig. \ref{Fig4}(b)). Unlike unconstrained matter for which
pairwise interactions suffice, for a chain molecule, it is necessary
to define the context of the object that is part of the chain.  This
is most easily carried out by defining a local Cartesian coordinate
system (see Fig. \ref{Fig6}) whose three axes are defined by the
tangent to the chain at that point, the normal, and the binormal which
is perpendicular to both the other two vectors. A
study\cite{HoangPNAS} of the experimentally determined native state
structures of proteins from the Protein Data Bank\cite{PDB} reveals
that there are clear amino acid aspecific geometrical constraints on
the relative orientation of the local coordinate systems due to
sterics and also associated with amino acids which form hydrogen bonds
with each other (see Fig. \ref{FigA3} in Appendix D).

Recently\cite{HoangPNAS}, we have carried out Monte Carlo
simulations of short {\em homopolymers}, chains made up of just
one type of amino acid,  subject to these geometrical constraints
and physically motivated interaction energies, a local bending
energy penalty, $e_R$, an overall hydrophobicity, $e_W$, and
effective hydrogen bond energies (see Appendix E for details about
the refined tube model and the simulations). The resulting phase
diagram and the associated structures for short homopolymers of
length $24$  are depicted in Fig. \ref{Fig7}. In keeping with the
behavior of the archetype tube discussed earlier, in the vicinity
of the swollen phase, one obtains distinct assembled tertiary
structures, quite akin to real protein structures, on making small
changes in the interaction parameters. The striking
similarity between the observed structures and real protein
structures suggests that our model captures the essential
ingredients responsible for the limited menu of protein native
structures.

The marginally compact phase has distinct structures including a
single helix, a bundle of two
helices, a helix formed by $\beta$-strands, a $\beta$-hairpin,
three-stranded $\beta$-sheets with two distinct topologies and a
$\beta$-barrel like conformation.  These structures are
the stable ground states in different parts of the phase diagram.
Furthermore, conformations such as the $\beta-\alpha-\beta$ motif are found
to be competitive local minima.
The specific structure
depends on the
precise values of the local radius of curvature penalty (a large penalty
forbids tight turns associated with helices resulting in an advantage for
sheet formation) and the strength of the hydrophobic interactions (a
stronger overall attraction leads to somewhat
more compact well-assembled tertiary structures).  The topology
of the phase diagram allows for the possibility of conformational
switching leading to the conversion of an $\alpha$-helix to a
$\beta$-topology on changing the hydrophobicity parameter
analogous to the influence of denaturants or alcohol in
experiments\cite{Switch}.

\section{Consequences of the protein-tube hypothesis}

We now turn to a study of some of the consequences of our
postulate that the tube is a useful paradigm for understanding
protein structures and behavior.
We will benchmark these against experimental evidence to assess
their validity.

\subsection{Energy landscape of proteins}

\begin{figure}
\centering \epsfig{width=8.0cm,file=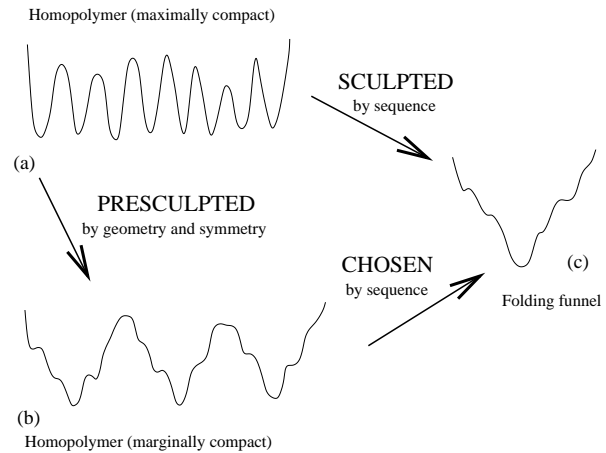} \caption{Simplified
one dimensional sketches of energy landscape. The quantity plotted
on the horizontal axis schematically represents a distance between
different conformations in the phase space and the barriers in the
plots indicate the energy needed by the chain in order to travel
between two neighboring local minima. (a) Rugged energy landscape
for a homopolymer chain with an attractive potential promoting
compaction as, e.g., in a string and beads model. There are many
distinct maximally compact ground state conformations with roughly
the same energy, separated by high energy barriers (the degeneracy
of ground state energies would be exact in the case of both
lattice models and off-lattice models with discontinuous
square-well potentials). (b) Pre-Sculpted energy landscape for a
homopolymer chain in the marginally compact phase. The number of
minima is greatly reduced and the width of their basin increased
by the introduction of geometrical constraints. (c) Funnel energy
landscape for a protein sequence. As folding proceeds from the top
to the bottom of the funnel, its width, a measure of the entropy
of the chain, decreases cooperatively with the energy gain. Such a
distinctive feature, crucial for fast and reproducible folding,
arises from careful sequence design in models whose homopolymer
energy landscape is similar to (a). In contrast, funnel-like
properties already result from considerations of geometry and
symmetry in the marginally compact phase (b), thereby making the
goals of the design procedure the relatively easy task of
stabilization of one of the pre-sculpted funnels followed by the
more refined task of fine-tuning the putative interactions of the
protein with other proteins and ligands.} \label{Fig8}
\end{figure}

There have been many previous studies of proteins from a physics point
of view\cite{Grosberg}. The standard approach is to assume an overall
attractive short range potential which serves to lead to a compact
conformation of the chain in its ground state.  In the absence of
amino acid specificity or when one deals with a homopolymer, there is
a huge number of highly degenerate ground states comprising all
maximally compact conformations with high barriers between them (see
Fig. \ref{Fig8}(a)). The ground state degeneracy and the height of the
barriers grow exponentially with the length of the homopolymer. The
role played by sequence heterogeneity is to break the degeneracy of
maximally compact conformations, leading to a unique ground state
conformation which, of course, depends on the amino acid
sequence. Yet, for a typical random sequence, the energy landscape is
still very rugged and is virtually the same as in Fig.\ref{Fig8}(a). A
model protein moving in such a rugged landscape can be subject to
trapping in local minima and may not be able to fold rapidly, so that
glassy behavior may ensue due to such trapping.  Bryngelson and
Wolynes\cite{Bryngelson} suggested that there is a principle of
minimal frustration at work for well-designed sequences in which there
is a nice fit between a given sequence and its native state structure
carving out a funnel-like landscape\cite{Funnel} which promotes rapid
folding and avoids the glassy behavior (Fig. \ref{Fig8}(c)).

Indeed, given a sequence of amino acids, with all the
attendant details of the side chains and the surrounding water,
one obtains a funnel-like landscape with the minimum corresponding
to its native state structure. Each protein is characterized by
its own landscape.  In this scenario, the protein sequence is
all-important and the protein folding problem, besides becoming
tremendously complex, needs to be attacked on a protein-by-protein
basis.

In contrast, our model calculations show that the large number of
common attributes of globular proteins\cite{BanavarRMP,Bernal}
reflect a deeper underlying unity in their behavior. At odds with
conventional belief, a consequence of our hypothesis is that the
gross features of the energy landscape of proteins result from the
amino acid aspecific common features of all proteins. This
landscape is {\em (pre)sculpted} by general considerations of
geometry and symmetry (Fig. \ref{Fig8}(b)). Our unified framework
suggests that the protein energy landscape ought to have thousands
of broad minima corresponding to putative native state structures.
The key point is that for each of these minima the desirable
funnel-like behavior is already achieved at the homopolymer level
{\em in the marginally compact part of the phase diagram} (see
Fig. \ref{Fig7}). The self-tuning of two key length scales, the
thickness of the tube and the interaction range, to be comparable
to each other and the interplay of the three energy scales,
hydrophobic, hydrogen bond, and bending energy, in such a way as
to stabilize marginally compact structures, also provide the close
cooperation between energy gain and entropy loss needed for the
sculpting of a funnelled energy landscape.

Recent work has shown that the rate of protein folding is not too
sensitive\cite{Baker,ZZ} to large changes in the amino acid
sequence\cite{ZZ,YY}, as long as the overall topology of the
folded structure is the same. Furthermore, mutational
studies\cite{Baker,TS1,TS2,TS3} have shown that, in the simplest
cases, the structures of the transition states are also similar in
proteins with similar native state structures.

Sequence design\cite{Design} would favor the appropriate native state
structure over the other putative ground states leading to a energy
landscape conducive for rapid and reproducible folding of that
particular protein.  Nature has a choice of 20 amino acids for the
design of protein sequences.  A pre-sculpted landscape greatly
facilitates the design process.  Indeed, within our model, we find
that a crude design scheme, which takes into account the hydrophobic
(propensity to be buried) and polar (desire to be exposed to the
water) character of the amino acids, is sufficient to carry out a
successful design of sequences with one or the other of the structures
shown in Fig. \ref{Fig7}.  The matching of the hydrophobic profile of
the designed sequence to the burial profile\cite{Match} (as measured
by the number of neighbors within the range of the hydrophobic
interaction) leads to the correct fold in a Monte Carlo simulation.
As examples, the sequence HPPHHPHHPPPPPPHHPHHPPPPP, with $e_R=0.3$
uniformly for all residues, $e_W=-0.4$ for contacts between H and H,
and $e_W=0$ for other contacts, has as its ground state the two-helix
bundle structure (Fig. \ref{Fig7}h) whereas HPHHHPPPPHHPPHHPPPPHHHPP
prefers the $\beta\alpha\beta$ motif (Fig. \ref{Fig7}j). It is
interesting to note that the $\beta\alpha\beta$ motif is only a local
minimum in the phase diagram of a homopolymer but is stabilized by the
designed sequence. Also, as is seen experimentally, many protein
sequences adopt the same native state conformation\cite{Manytoone}.
Once a sequence has selected its native state structure, it is able to
tolerate a significant degree of mutability except at certain key
locations\cite{TS1,TS2,Design,TSth}.  Furthermore, multiple protein
functionalities can arise within the context of a single
fold\cite{HolmSander}.

One of the successful methods of protein structure prediction is
based on threading\cite{Thornton}.  The basic idea is entirely consistent with
our findings -- one uses pieces of native state structures of
longer proteins as  possible candidate structures of a shorter
protein -- the technique is simpler because instead of
determining the structure from ab-initio calculations, one
merely has to select from among the putative native state
structures.   The documented success of the threading method
confirms that each protein does not fashion its own
native state structure but merely selects from the menu of
pre-determined folds.

\subsection{Amyloid phase of proteins}

A range of human diseases such as Alzheimer's, spongiform
encephalopathies and light-chain amyloidosis lead to degenerative
conditions and involve the deposition of plaque-like material in
tissue arising from the aggregation of
proteins\cite{Kelly,Ddisease,Dobson1,Prusiner}.
In the case of prions \cite{Prusiner},
one observes a transition from $\alpha$ to $\beta$ rich structures
which favors aggregation and causes
bovine spongiform encephalopathy (BSE) disease.
It has been argued\cite{Aggeli} that the  formation of amyloid fibrils occurs
in a hierarchical way starting from a chiral $\beta$-strand.
The resulting structures arise from a competition between the free energy gain
from the aggregation and the elastic energy cost of the distortion.
A  variety of proteins not involved in these diseases also form
aggregates very similar to those implicated in the
diseased state\cite{Ddisease,Dobson1}.   This
suggests\cite{Ddisease} that the
tendency for proteins to aggregate is a generic property
of polypeptide chains with the specific sequence of amino acids
playing at best a secondary role.   Can one understand this
general tendency of proteins to form amyloids within our
framework?

\begin{figure}
\centering \epsfig{width=3.2in,file=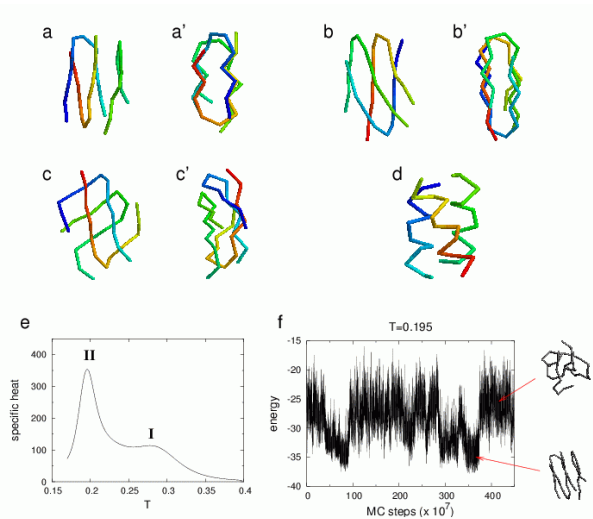} \caption{(Color online) Aggregated
structures formed by three chains of length 12. We show the lowest
energy conformations obtained in long simulations for three 12-residue
chains confined within a cubic box of side $L=80$ \AA\/ at $T=0.19$ (a),
$T=0.18$ (b) and $T=0.16$ (c).  The conformations shown in (a'), (b')
and (c') are the same as in (a), (b), and (c) respectively, but viewed
from a different angle.  The parameters used in the model are
$e_W=-0.08$ and $e_R=0.2$ which correspond to having a single helix
ground state in the case of a single chain. The simulations start with
random extended conformations for all chains and are carried out with
pivot and crank-shaft moves that are accepted or rejected based on the
Metropolis criterion. Moves that bring the residues out of the box are
not allowed. The bundle of 3 helices (d) is a putative ground state of
the system and was obtained in a simulation at a very low temperature
($T=0.05$) starting with isolated single helices. This conformation
has the lowest energy among those shown but is not the equilibrium
conformation at intermediate temperatures. Indeed, a simulation run at
$T=0.18$ starting with conformation (d) leads to the helix bundle
being converted into the $\beta$-helix-like conformation shown in (b)
which is the dominant equilibrium conformation at this temperature.
e) The specific heat as function of temperature for the system of
three 12-residue peptides. The data shown were obtained using the
weighted histogram technique\cite{FerSwen} based on long equilibrium
simulations at various temperatures between 0.16 and 4. The small
shoulder (I) corresponds to a condensation of separated peptides into
a disordered globule. The large peak (II) corresponds to a transition
from disordered globule to the $\beta$-helix-like phase. f) The
energy as a function of time (in Monte Carlo steps) during a long
simulation at a temperature corresponding to the maximum of the
specific heat, $T=0.195$. The simulation shows several transitions
between the disordered globule phase and the $\beta$-helix-like
phase.} \label{Fig9}
\end{figure}

Let us recall the semi-crystalline polymer phase which one
obtains when the tube is sufficiently long (or
when there are many interacting tubes) and is
subject to attractive interactions leading to
compaction.  In this phase, the tube is stretched out
locally with nearby sections parallel to each other (or has the
tubes stacked parallel to each other in a periodic arrangement) and does not
have the richness we associate with protein native state
structures.   Returning to the protein, one may ask
whether there are structures which are the analogs of those found
in the semi-crystalline phase.

In order to assess the role played by the interaction between multiple
short proteins, let us first consider our model homopolymer chain made
up of 36 identical amino acids in the marginally compact phase of the
refined tube model (see Section IV and Appendix E) with the
hydrophobic parameter and the local bending energy penalty chosen so
that the ground state is a single long helix.

\begin{figure}
\centering
\epsfig{width=8.0cm,file=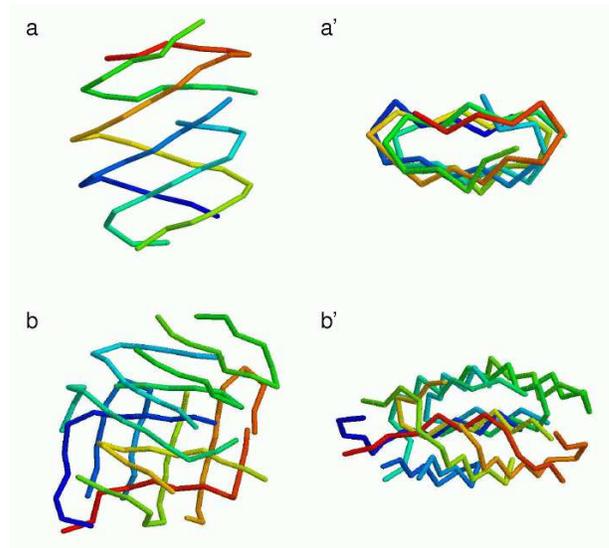}
\caption{(Color online) Aggregated structures formed by five and ten chains of length
  12 with $e_W=-0.08$, $e_R=0.2$. We show the lowest energy
  conformations obtained in long simulations for five chains at
  $T=0.18$ (a), and for ten chains at $T=0.2$ (b). The five chain
  system is confined within a cubic box of side $L=80$ \AA\/ whereas the
  ten chain system is confined within a cubic box of side $L=100$ \AA.
  The conformations shown in (a') and (b') are the same as those in
  (a) and (b) but viewed from a different angle.}
\label{Fig10}
\end{figure}

On making two incisions in the chain to create three distinct
chains each containing 12 amino acids, the ground state of the
system appears to be a bundle of three helices (see Fig.
\ref{Fig9}(d)). This helix structure however is stable only at
very low temperatures. At intermediate temperatures, close to but
lower than the temperature of the specific heat peak, it is
destabilized in favor of aggregated $\beta$-helices (Figures
\ref{Fig9}(a) and \ref{Fig9}(b)) or sandwiches of $\beta$-sheets
(Fig. \ref{Fig9}(c)), due to entropic effects. Cutting a single
chain into parts increases the entropy of the system. Unbonded
chains are more flexible and this promotes the formation of
interchain hydrogen bonds. The $\beta$-sheet structures also show
an increased flexibility comparing to the helix bundle, and they
have better kinetic accessibility from a disordered globule. While
the appearance of $\beta$-sheet conformations in the case of three
chains seems to have an entropic origin, it seems likely that the
ground state of a system of multiple chains does in fact consist
of aggregated $\beta$-sheets. Indeed simulations of 5 or 10 chains
have shown that $\beta$-structures are the most likely choice (see
Fig. \ref{Fig10}).

The formation of $\beta$-sheet structured protein aggregates is
favoured with respect to other possible aggregates such as helix
bundles (which we actually detect in our simulations, see
Fig. \ref{Fig9}(d)). In the latter case hydrogen bonds are saturated
within a single helix so that aggregation is driven exclusively by the
effective hydrophobic attraction between different helices.  On the
other hand, for structures such as those shown in
Fig. \ref{Fig9}~(a-c) hydrogen bonds are formed between different
chains and several $\beta$-strands are left unsaturated at both
``ends'' of the aggregate, which can then readily grow by
hydrogen-bonding to other chains.

\begin{figure}
\centering
\epsfig{width=8.0cm,file=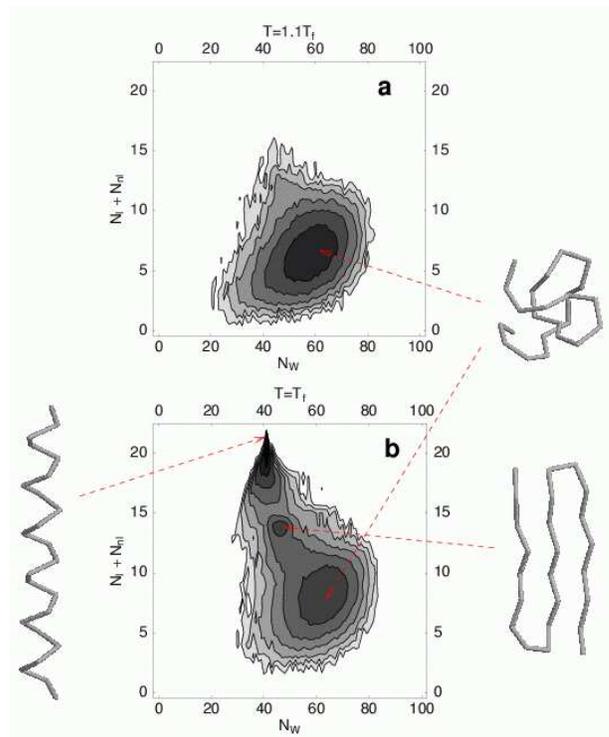}
\caption{(Color online) Contour plots of the effective free energy (a) at high
temperature ($T = 0.22$) and (b) at the folding transition temperature
$T_f=0.2$ for a single 24-residue homopolymeric chain, with
$e_W=-0.08$, $e_R=0.3$. The effective free energy, defined as
$F(N_l+N_{nl},N_W)=-\ln P(N_l+N_{nl},N_W)$, is obtained as a function
of the total number of hydrogen bonds $N_l+N_{nl}$ and the total
number of hydrophobic contacts $N_W$ from the histogram
$P(N_l+N_{nl},N_W)$ collected in equilibrium Monte-Carlo simulations
at constant temperature. The spacing between consecutive levels in
each contour plot is $1$ and corresponds to a free energy difference
of $k_B\tilde{T}$, where $\tilde{T}$ is the temperature in physical
units. The darker the color, the lower the free energy value.  There
is just one free energy minimum corresponding to the denatured state
at a temperature higher than the folding transition temperature (Panel
(a)) whereas one can discern the existence of three distinct minima at
the folding transition temperature (Panel (b)).  Typical conformations
from each of the minima are shown in the figure.}
\label{Fig11}
\end{figure}

The refined tube model can be used to explore the free energy
landscape of a homopolymer chain in the vicinity of its folding
transition temperature, operationally defined as the specific heat
peak temperature.  (Of course, there is no real phase transition for
finite size systems such as proteins.)  Fig. \ref{Fig11}(a) is a
contour plot of the free energy at a temperature higher than the
folding transition temperature for the parameter values $e_W=-0.08$
and $e_R=0.3$ for which the ground state is an $\alpha$-helix. The
free energy landscape has just one minimum corresponding to the
denatured phase whose typical conformations are still somewhat
compact.  The contour plot at the folding transition temperature
(Fig. \ref{Fig11}(b)) has three local minima corresponding to an
$\alpha$-helix, a three-stranded $\beta$-sheet and the denatured
state. At lower temperatures, the $\alpha$-helix is increasingly
favored and the $\beta$-sheet is never the global free energy minimum.

That a $\beta$-sheet structure is a significant competitor with a
large basin of attraction in a region where the stable phase is a
helix (see Fig. \ref{Fig11}) reinforces the possibility that the
interaction between several proteins could stabilize the formation of
extended hydrogen bonded $\beta$-sheets via the aggregation of
individual chains (see \cite{Gruebele} for experimental evidence that
the increased propensity for extended single chain
$\beta$-conformations as the temperature is increased could indeed
drive the formation of $\beta$-aggregates). These kinds of structures,
which resemble the basic structures associated with amyloid fibrils,
thus seem to belong to the general class of pre-determined folds, but
this time for multiple proteins, and ought to be seen ubiquitously in
generic proteins\cite{Ddisease,Dobson1}.  This suggests that the key
to the prevention of such aggregates is the stabilization of helices
in such proteins and evolutionary mechanisms such as proteasomes,
molecular chaperones\cite{Chap} and ubiquitination
enzymes\cite{Ddisease,Dobson1}.

Our results show the generic tendency for multiple chains of
amino acids to form aggregated amyloids rather than maintain their
protein-like shape.  Interestingly, nature has, on suitable
occasions, thwarted the tendency of a single long chain to form
amyloid by dividing the protein into substantially independent
domains which fold autonomously and are then assembled together.
This suggests that the variety of protein folds increases with
length up to a certain point at which they are supplanted by the
formation of domains or amyloids.

In a recent paper, Fandrich and Dobson\cite{Fandrich} suggested that "amyloid
formation and protein folding represent two fundamentally
different ways of organizing polypeptides into ordered
conformations.  Protein folding depends critically on the presence
of distinctive side chain sequences and produces a unique
globular fold. By contrast, .... amyloid formation arises
primarily from main chain interactions that are, in some
environments, overruled by specific side chain contacts." Our
results are in complete accord with the suggestion that amyloid structures
may arise from the generic
properties of the proteins with the details of the amino
acid side chains playing a secondary role.  However, our work
suggests that instead of an "inverse side chain effect in amyloid
structure formation"\cite{Fandrich}, there is a unifying theme in the behavior
of proteins.  Just as the class of cross-linked $\beta$-structures
are determined from geometrical considerations,
the menu of protein native state structures is also
determined by the common attributes of globular proteins: the
inherent anisotropy associated with a tube and the geometrical
constraints imposed by hydrogen bonds and steric considerations.

\subsection{Natural selection and protein interactions}

Traditionally, the framework of evolution in life works through two
aspects of organization called the genotype and the phenotype. The
genotype is the heritable information encoded in the DNA, which is
translated through the RNA molecules into proteins.  The phenotype is
valuable for adaptation and at the molecular level plays a key role in
natural selection.  One conventionally assumes that there is a
selection of phenotypes which leads to an enhancement in the numbers
of the genotype.  Furthermore, mutations of the genotype lead to the
possibility of new phenotypes.

Let us consider the situation at two levels: the sequence level
(which is the genotype because it is a direct translation from the
evolving DNA molecules) and the structure level, which we can
think of as the phenotype.  As pointed out by
Maynard-Smith\cite{Maynard}, as the sequence undergoes mutation,
there must be a continuous network that the mutated sequences can
traverse without passing through any intermediaries that are
non-functioning.  Thus, one seeks a connected network in sequence
space for evolution by natural selection to occur.  There is
considerable evidence, accumulated since the pioneering suggestion
of Kimura\cite{Kimura} and King and Jukes\cite{King}, that much of
evolution is neutral.  The experimental data strongly supports the
view that the "random fixation of selectively neutral or very
slightly deleterious mutants occur far more frequently in
evolution than selective substitution of definitely advantageous
mutants."\cite{Kimura1}.  Also "those mutant substitutions that
disrupt less the existing structure and function of a molecule
(conservative substitutions) occur more frequently in evolution
than more disruptive ones"\cite{Kimura1}. Thus while one has a
``random walk'' in sequence space that forms a connected network,
there is no similar continuous variation in structure
space\cite{Koonin,Tiana}.

These facts are in accord with our result of a pre-sculpted
energy landscape that is shared by all proteins and has thousands of
local minima corresponding to putative native state structures -- not
too few because that would not lead to sufficient diversity and not
too many because that would lead to too rugged a landscape with
little hope that a protein could fold reproducibly and rapidly into
its native state structure.  Indeed, many proteins share the same
native state fold and often the mutation of one amino acid into
another does not lead to radical changes in the native state
structure underscoring the fact that it is not the details of the
amino acid side chains that sculpt the energy landscape but
rather some overarching features of symmetry and geometry that are
common to all proteins.  In this respect, the phase of matter that
comprises the native state structures is one that is possibly
determined by physical law rather than by the plethora of microscopic
details in analogy with the limited menu of possible crystal
structures.

Anfinsen\cite{Anfinsen} wrote in 1972, ``Biological function appears
to be more a correlate of macromolecular geometry than of chemical
detail.'' There has been much recent progress in extracting
information on biological function and protein
interactions\cite{vonMering} from the structure of proteins and the
complexes they form\cite{ThorntonNSB}. A protein structure chosen
from the predetermined menu of folds contains information on the
topology of the folded state.  Additionally, one can glean
information on the nature of the exposed surface, crystal packing,
and the existence of clefts or other geometrical features (which are
often the active sites of enzymes).  The picture is completed by
knowledge of the sequence of amino acids that folds into the
structure using which one can infer the amino acid composition of the
exposed surfaces, the location of mutants and conserved residues and
evolutionary relationships. For some structural families, function is
highly conserved, whereas for others, one can use the types of
information described above to guess the function\cite{Orengo}.

Biological reactions are accelerated by factors more than a
billion by enzymatic proteins. Enzymes not only provide for great
catalytic efficiency but are also extremely specific in their
function.  The principal mechanism\cite{karplusscience2004}
underlying the tremendous enhancement of the reaction rate by the
enzymes is the lowering of the free energy of the transition state
of the reaction through their specific binding to the substrate or
the reactant(s).  In its native state, an enzyme adopts a
structure chosen from the menu of pre-determined folds.
Strikingly, only a small part of this structure is important for
the enzymatic action.  Generally, there are a few amino acids,
associated with the active site, which are responsible for the
catalytic activity. In close proximity, one also finds the
substrate binding site which provides the specificity, often
through the classic lock and key mechanism.

An illustration of enzymatic action and the role of molecular
evolution is provided by the protease family of proteins. In a
living cell, there is turnover of proteins with new proteins being
continually synthesized along with the degradation of existing
proteins.  Proteins responsible for degradation through the
hydrolysis of peptide bonds are called proteases.   Under
physiological conditions, peptide bonds are stable for a period of
around a hundred years.  The proteases are able to enhance the
degradation rate selectively by factors of around a billion. There
are several classes of proteases including serine proteases (such as
chymotrypsin, a digestive enzyme) with a very reactive serine
residue, cysteine proteases (such as papain, which is a digestive
enzyme derived from papaya) with cysteine playing the role of serine,
aspartyl proteases (such as renin which controls blood pressure)
which employs a pair of aspartate groups and metalloproteases (such
as collagenase responsible for collagen degradation in
osteo-arthritic cartilage), which use a bound metal ion such as zinc
to accelerate the hydrolysis.

In serine proteases, the catalytic triad comprises three amino acids,
serine, histidine and aspartate, bound to each other through hydrogen
bonds, whose presence leads to the proton being moved away from the
serine and the creation of a reactive alkoxide ion.  The same triad
is implicated in all serine proteases.  Indeed, an example of
convergent evolution is provided by subtilisin (an enzyme that
resembles chymotrypsin in its action and is made by certain soil
bacteria) and its family members, which possess the catalytic triad
but have a quite different structure from chymotrypsin. Here Nature
uses different folds from the pre-sculpted energy landscape
which, on appropriate sequence design, have the same catalytic triad
and perform similar tasks.

The limited menu of possible protein folds provides a marvellous
opportunity for divergent evolution.  This corresponds to proteins
whose native state structure and the catalytic triad are the same
but with distinct differences in the nature of the binding site.
The binding site in chymotrypsin is adjacent to the active site
and is a hydrophobic cavity which facilitates hydrolysis of the
peptide bonds on the carboxyl side of aromatic or large
hydrophobic amino acids such as Trp, Tyr, Phe, Met and Leu.
Relatively small changes in the amino acid sequence, which
maintain both the native state structure and the active triad lead
to other proteins such as trypsin (a digestive protein made in the
pancreas which cleaves after positively charged amino acids lysine
and arginine due to a change of one of the hydrophobic amino acids
in the binding cavity to a negatively charged aspartic acid),
elastase (a protein made both in the pancreas and by white blood
cells in which two glycines in the binding cavity are replaced by
much larger amino acids valine and threonine allowing the enzyme
to specifically target elastin, which is an important building
block of blood vessel walls and ligaments -- elastase is able to
cleave proteins after a glycine and alanine because of the small
size of the binding cavity), thrombin (a larger enzyme, the
tail-end of which bears a significant similarity to the sequence
of amino acids of chymotrypsin and trypsin and cleaves proteins
only at arginine-glycine linkages; thrombin is a complex
regulatory protease which converts a usually soluble blood protein
fibrinogen into the insoluble fibrin causing a blood clot and the
cessation of bleeding), plasmin (an enzyme which cleaves proteins
after lysine and arginine and dissolves blood clots), cocoonase
(which also cleaves after lysine and arginine in the silk strands
of the cocoon after the transformation of a caterpillar into a
silk moth) and acrosin (an enzyme which plays a pivotal role in
fertilization by creating a hole in the protective sheath around
the egg and allowing sperm-egg contact).

As we have seen, evolution along with natural selection allow
Nature to use variations on the same theme facilitated by the rich
repertory of amino acids to create enzymes that are able to
catalyze a remarkable array of diverse and complex tasks in the
living cell. The key point, of course, is that in order for
molecular evolution to work in this manner, one needs the constant
backdrop of folds shaped not by the sequence but determined by
physical law.  Were the folds not immutable and themselves subject
to Darwinian evolution, the possibility of creating so many subtle
and wonderful variations on the same theme would not exist. The
pre-sculpted landscape is the crucial feature that leads to a
predetermined menu of immutable folds.

It is known that key functional sites exhibit a high degree of
conservation\cite{Lichtarge}.  Interestingly, co-evolutionary analysis has
been useful in identifying protein-protein interactions\cite{GohBickel}.
Structural similarity, independent of evolutionary homology, can be the
key reason why proteins with different folds share some commonality in
enzymatic activity or ligand binding\cite{Campbell}.  Conversely, there
are protein structures such as the TIM barrel\cite{Nagano} which are very
versatile and are able to house proteins that are able to carry out
multiple functionalities.  Even though the proteins are able to perform
diverse catalytic tasks, Nagano et al.\cite{Nagano} find that the active
site is generally found at the C-terminal end of the barrel sheets and
that there are ``striking structural superpositions" of the metal-ligating
and catalytic residues.

Nooren and Thornton\cite{Nooren} have pointed out that ``The structure and
affinity of a PPI (protein-protein interaction) is tuned to its biological
function and the physiological environment and control mechanism.  PPIs
presumably evolve to optimize `functional' efficacy.  This does not
necessarily involve strong interactions.  Clearly, weak transient
interactions that are efficiently controlled are also very important in
cellular processes.''

There are several attractive features of the picture we have developed
based on the tube-protein hypothesis. First, protein structures lie in the
vicinity of a phase transition to the swollen phase which confers on them
exquisite sensitivity, especially in the exposed parts of the structure,
to the effects of other proteins and ligands.  The flexibility of
different parts of the protein depend on the amount of constraints placed
of them from the rest of the protein\cite{Thorpe123}.  From this point of
view, it is easy to understand how loops, which are not often stabilized
by backbone hydrogen bonds can play a key role in protein functionality.

It is useful to reconsider how nature uses the variety of amino
acids for sequence design.  The existence of a pre-sculpted energy
landscape with broad minima corresponding to the putative native
state structures and the existence of neutral evolution
demonstrates that the design of sequences that fit a given
structure is relatively easy leading to many sequences that can
fold into a given structure.  This freedom facilitates the
accomplishment of the next level task of evolution through natural
selection: the design of optimal sequences, which not only fold
into the desired native state structure, but also are fit in the
environment of other proteins. A useful protein is one that can
interact with other proteins in a synergistic manner and at the
same time is not subject to the tendency to aggregate into the
harmful amyloid form.  This suggests that protein engineering
studies aimed at improving enzymatic function ought to be carried
out in a two step manner: first, the family of sequences that fold
into a desired target structure need to be selected and a finer
design needs to be carried out in the context of the substrates
and the other proteins that the target protein interacts with.
Unlike the generality of geometry and symmetry that leads to the
menu of native state folds, what we have here is a problem of
chemistry acting within the fixed background of the physically
determined structures. These considerations suggest that, when the
information becomes available, protein-protein interaction
networks\cite{PPINature} can be fruitfully viewed not only as the
interactions between proteins but also as the interactions between
the structures that house them.

The characteristics required for protein native state structures to be
targets of an evolutionary process are stability and diversity.
Stability is needed because one would not want to mutate away a DNA
molecule able to code for a useful protein, and diversity, in order to
allow evolution to build complex and versatile forms.  The mechanism for
natural selection arises naturally in this context -- DNA molecules that
code for amino acid sequences that fit well into one of these
predetermined folds and have useful functionality thrive at the expense of
molecules that create sequences that are not useful. Indeed, in this
picture, sequences and functionality evolve in order to fit within the
constraints of these folds, which, in turn, are immutable and determined
by physical law.

\section{The denatured state of proteins}

Progress occurs in science through the use of constructive
hypotheses with a careful assessment of their consequences.
Experiments not only provide valuable hints for selecting between
competing hypotheses but are also the ultimate test of a given
hypothesis. There are strong hints from protein experiments that the
protein-tube hypothesis is valid.  It provides a
unification of the various aspects of all proteins: one obtains a
pre-sculpted energy landscape with relatively few folds,
one can rationalize how a protein might fold in a cooperative
manner into its native state conformation, there is the
possibility of straightforward design of optimal sequences that
fit into a desired structure, the structures are in a marginally
compact phase in the vicinity of a phase transition and have the
flexibility needed for biological function, and one can
understand the formation of amyloids and the role played by the
protein structures as a molecular basis for natural selection.

Protein sequence design provides an optimal fit of the sequence
with one among the menu of pre-sculpted conformations.  The
question arises of course as to how a given sequence is able to
reach its native state conformation or its home starting from its
denatured conformation.  The answer to this question entails the
understanding of its denatured
state\cite{polyproline,Dillshortle,PSR,Plaxco,Rosedenatured,Shortle,Gunsteren,Pande}.
Unlike the native state which is a somewhat tightly bound set of
marginally compact conformations, one envisions the denatured
state as an ensemble of somewhat open conformations that the
protein adopts when it is not under physiological conditions.

While one may naively think that the denatured state is devoid of any
interesting features, recent work has underscored the possibility
that the number of accessible conformations is severely reduced
compared to a random
chain\cite{polyproline,Dillshortle,PSR,Rosedenatured,Shortle,Gunsteren,Pande}
leading to biases in the chain
direction that persist over the entire length of the protein\cite{Shortle}.
Indeed, Shortle\cite{Shortle} has argued that ``long-range
structure, which cannot be removed by strongly denaturing
conditions, could arise predominantly from local steric
hindrance.''   He goes on to state that ``not only does the
ribosome determine the primary structure of each protein it makes,
it also establishes the topological space in which that protein
chain will be confined for the rest of its existence.''

We build on these insights and the presumed validity of our
protein-tube hypothesis by making a second
hypothesis that just as there is a
one-way correspondence between a sequence and its native state
structure, there could exist a similar correspondence between the
sequence and its denatured state. In this view, the denatured
state can thought of as an address of the native
state conformation and lies within its basin of attraction.

Unlike the native state, the denatured state has a larger entropy and
comprises somewhat open conformations. Because of this, water plays a
quite crucial role in the denatured state. Both the above factors lead
to local interactions\cite{Secondary,contactorder} playing a more
important role than non-local interactions in the denatured state.  As
can be seen from Fig. \ref{FigA3}(a), the local bending energy term is
amino acid specific. In addition, in the spirit of the tube model, one
might ask whether there are extra geometrical constraints between the
local frames of reference (see Fig. \ref{Fig6}) of neighboring amino
acids along the chain. (As discussed earlier, at the non-local level,
hydrogen bonds linking different parts of the chain do place
geometrical constraints on the reference frames associated with these
locations.)  Physically, such correlations arise from the fact that in
addition to the $C^{\alpha}$ atom that we have considered as a
surrogate for the amino acid, all amino acids but glycine have a
$C^{\beta}$ atom to which the side chain is attached.  In a chain of
coins (see Section II), this corresponds to breaking the symmetry in
the plane of the coin. Thus one would quite generally expect that side
chain interactions would lead to correlations between the local
coordinate frames of nearby amino acids along the
sequence\cite{Scheragadiff}.  Remarkably, the local steric
constraints\cite{Rama} and the hydrogen bonds\cite{Pauling1,Pauling2}
act in concert and both promote helices and sheets in the native
state.

One can ask what the effects of such a local interaction are in the
absence of any non-local interaction promoting the compaction of the
chain.  Let us first consider a homopolymer made of just one kind of
amino acid.  A simple chain molecule with a local bending constraint
leads to a tangent-tangent correlation, $\langle \hat t_i \cdot \hat
t_{i+n}\rangle$ (see Fig. \ref{Fig6} for the definition of the tangent
vector), that decays exponentially in $n$.  Adding a local
binormal-binormal interaction term leads quite generally (see the
example in Appendix F) to the tangent-tangent correlation decaying
exponentially with sequence separation but being modulated with
oscillatory behavior.  This generic behavior underscores the fact that
the class of denatured conformations are not merely featureless but
rather already have short-range structure built into them. Indeed,
there is a clear reduction in the entropy, due to the short-range
binormal-binormal interaction, which is reflected in the oscillations.

The situation is vastly more interesting when one considers a specific
sequence of amino acids in its denatured state.  It is clear that one
ought to have amino acid specific correlations between neighboring
coordinate frames which reflect the nature and size of the side
chains.  An amino acid like proline with its cumbersome side chain
configuration can lead to strong constraints in its vicinity whereas
glycine which lacks the $C_\beta$ atom and a side chain can provide
great flexibility\cite{Joker} and act as a joker in a card game.  Even
if such correlations reflect small but systematic deviations from the
average behavior\cite{Kiefhaber}, these can build up in a very
specific way along the sequence leading to a clear imprinting of the
native state conformation even in the denatured state.  In this
context, it is interesting to note that Shortle\cite{Shortle} has
shown that "denaturation by at least three different agents --
truncation, urea and acid -- gives rise to essentially the same
persistent native-state like topology".  Furthermore, the alteration
of the denatured state by even a single mutation\cite{Mutden} provides
further evidence for the structure inherent in the denatured state.

We have shown that the menu of native state
structures is determined from generic considerations.  Sequence
specificity is key in determining whether a given sequence fits
particularly well into one of these conformations.  Because the
menu is large (thousands of conformations), one has diversity.
However, because the menu is not too large, a well-designed
sequence is able to fold rapidly into its native state
conformation.  Our hypothesis is that  local sequence-specific
interactions alone lead to a denatured state which is a
reflection of the native state.  The denatured state lies in the
basin of attraction of the native state and the folding process
simply entails the action of the appropriate non-local
interactions in leading to the protein adopting
the native state conformation.

The situation is somewhat
reminiscent of a content-addressable memory\cite{Hopfield} in which partial
information is converted by the brain to recover the complete
information.  Such content addressable memories\cite{Hopfield} as well as
the energy landscape\cite{Anderson} suitable for prebiotic
evolution\cite{Eigen} have been
modeled through spin glasses\cite{Spinglass}.  The energy landscape of spin
glasses is also characterized by diversity and stability arising
from randomness and frustration which
is quite distinct from the
the physical mechanisms of short tubes in the marginally compact
phase.  In conventional spin glasses, randomness, which plays a
role somewhat similar to amino acid specific interactions in
proteins\cite{Spinglassprot},
through frustration
sculpts an energy landscape with many local minima.  Indeed, a
non-random exchange interaction between spins would lead to periodic order
with much simpler behavior.  In spin glasses, starting from a random
spin configuration, it is hard to reach a specific local minimum unless
the exchange constants are tuned in a clever way as in a content
addressable memory.  The landscape is not invariant on changing
the exchange interactions and can be fashioned at will.
For proteins, on the other hand,  our analysis shows that a rich
landscape is obtained even in the absence of any sequence
heterogeneity and the nature of the ground states is determined
by geometry and symmetry and is therefore immutable\cite{Denton}.

An interesting consequence of the type of denatured state described
above along with the existence of the pre-sculpted landscape is the
possibility of disordered proteins\cite{Disordered} -- sequences that
are in temporally fluctuating denatured form but which fold in the
presence of distinct substrates to carry out vital multiple
functionalities.  In our picture, these sequences need appropriate
stabilizing influences to fold. In the absence of these influences
(substrates), the protein is denatured and is located, colloquially,
on the fence between different native state structures.  Given that
finite size effects are severe for proteins, the presence of different
substrates (leading to different boundary conditions) would not only
favor one competing structure over the others but also result in
folding to that structure.  The simultaneous existence of the distinct
folds in the energy landscape allows the protein to choose from among
them depending on the precise nature of the stabilizing influence.

\section{Summary and perspective}

Symmetry and geometry place strong constraints on the types of
infinite sized crystal structures and there are exactly 230 distinct
space groups in 3 dimensions\cite{Yale}. Proteins are finite sized
objects. Our analysis demonstrates that the same kind of symmetry and
geometrical considerations lead to a finite number of protein folds.
This number grows with the size of the protein but is limited by the
fact that proteins beyond a characteristic length either form
autonomous domains or amyloids.  Unlike the crystalline state of
matter, proteins are characterized by an inherent anisotropy due to
their tube-like character. A given crystalline structure transcends
the material that is housed in it -- common salt adopts the
face-centered-cubic lattice structure as also the well-packed
cannonballs of Kepler\cite{Szpiro}.  Likewise, different sequences of
proteins can be housed in the same protein fold and yet be able to
perform different functionalities\cite{HolmSander}.  Protein
structures are modular in form being simple assemblages of helices and
strands connected by tight turns.

The unified picture leads to a single free energy landscape with
two distinct classes of structures. The amyloid phase is dominated
by $\beta$-strands linked to each other in a variety of forms
whereas the native state structure menu is an assembly of
$\alpha$-helices and $\beta$-structures. Nature has exploited
these native state structures in the context of the work horse
molecules of life.  The selection mechanism for genetic evolution
at the molecular level lies in the ability of the protein encoded
by the gene to fold well into one of the predetermined folds and
have useful function.  Unfortunately, however, the proximity of
this beautiful phase to the generic amyloid phase underscores how
life can easily malfunction as soon as aggregational tendencies of
proteins come to the fore.  One cannot but marvel at the
robustness of life.

An imperfect analogy to the protein problem is a township
consisting of around a thousand houses (protein structures),
each with its own
distinctive style (topology), determined by geometry and symmetry.
The form of a house (structure) is the basis of useful functionality.
A person (protein) whose tastes (sequence) are (is) especially matched to a given
style of house (native state structure) would choose to live in it.
Of course, many
people (proteins) with similar though not identical tastes
(sequences) might choose the
same style of house (native state structure).   If a person were to
arrive in this town, how would she/he know which house to move into?
 One way would be to explore all the house styles until the dream
house is identified.  A vastly more efficient situation would occur
if the person arrives at the township in the vicinity of the house
that she/he will eventually occupy.   This would require that
the location of the starting
point (the denatured conformation) is encoded by the tastes of
the person (the sequence) and is within the basin of attraction
of her/his dream home (native state structure). This, as yet
unproven, scenario
would greatly facilitate the folding of a protein into its native
state structure accounting for its ``surprising
simplicity''\cite{Baker}.

The protein problem, which lies at the intersection of many
disciplines, is highly complex.    Evolution complicates the
situation even further.  Human design allows for an engineer
to devise entirely new ways of accomplishing certain tasks -- a
classic example is the replacement of vacuum tubes with
semiconductor transistors.  Nature does not have this luxury in
evolutionary design.  Nature takes what she has, tinkers with it
and builds on it.  Thus the notion of optimal design is not
particularly relevant and the future is very strongly
correlated with the present and the past.  A slightly different
turn of events could have lead to conspicuously different life forms.
This picture of Nature muddling along through evolution combined
with the inherent complexity of proteins
makes the problem very daunting.
Yet, within this complexity, there is a stunning simplicity
provided by the fixed backdrop of the protein folds
determined by physical law in the
context of which sequences and functionalities are shaped by
evolution.

\begin{figure}
\centering
\hfill
\subfigure[]
{\includegraphics[height=3.5cm]{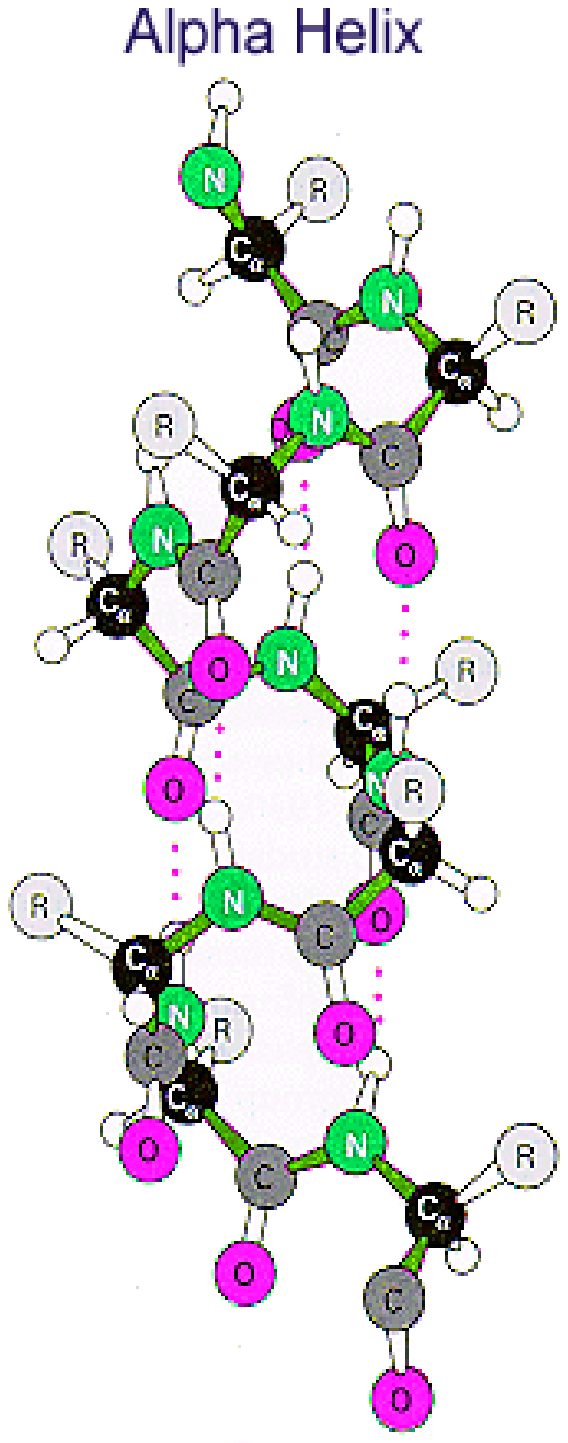}}
\hfill
\subfigure[]
{\includegraphics[height=3.5cm]{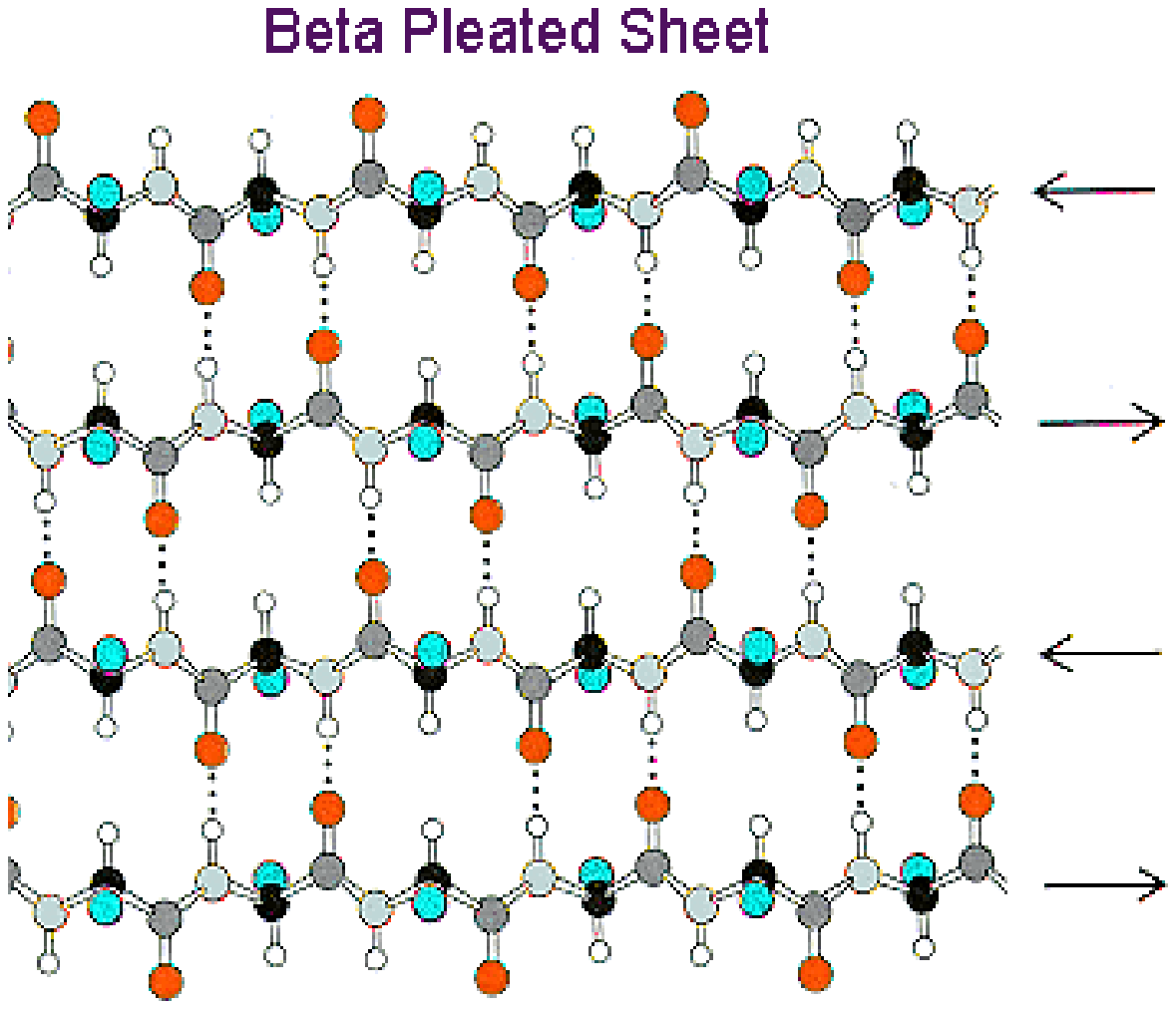}}
\hfill
\\
\hfill \subfigure[] {\includegraphics[height=3.5cm]{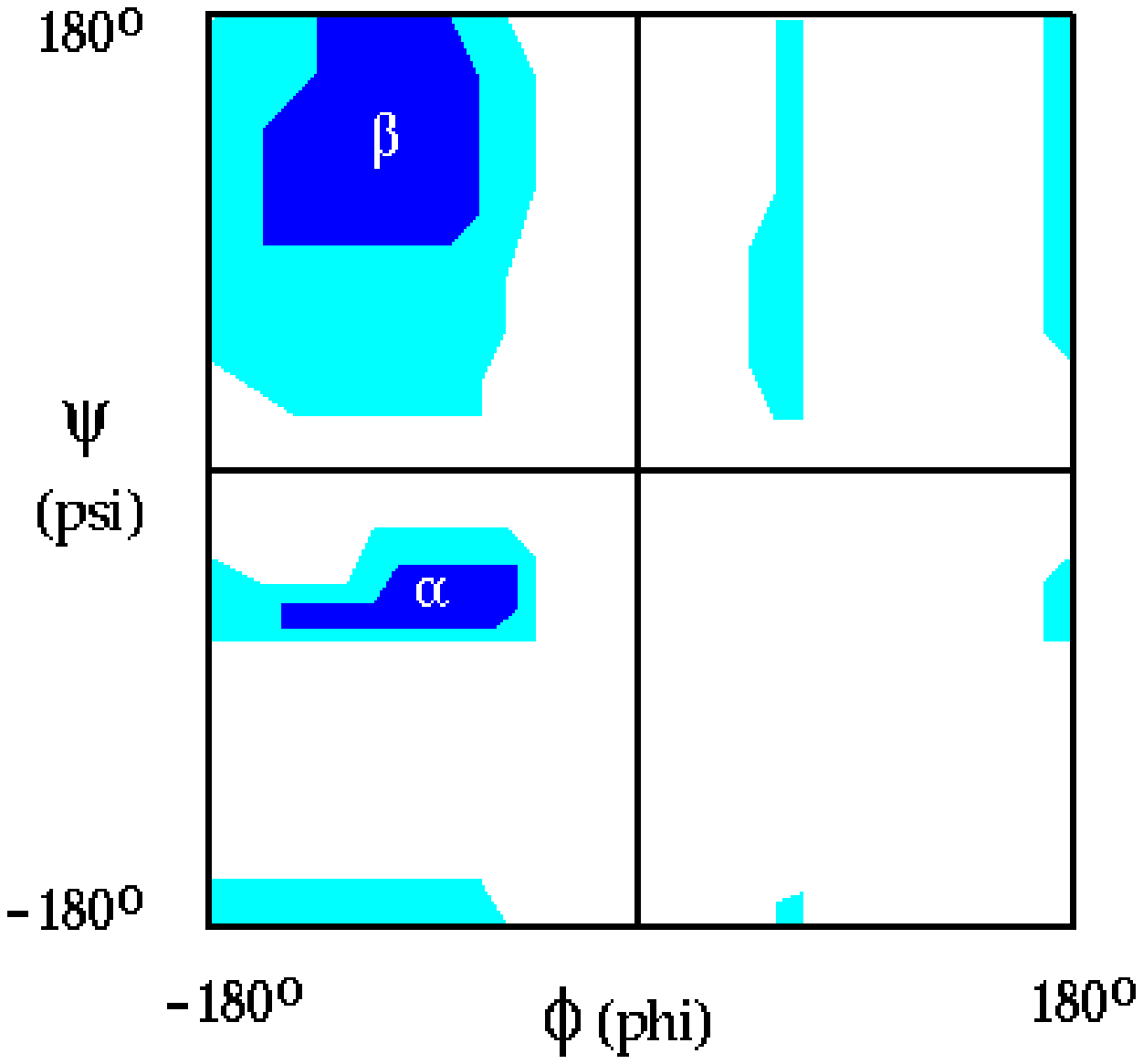}}
\hfill \subfigure[] {\includegraphics[width=3.5cm]{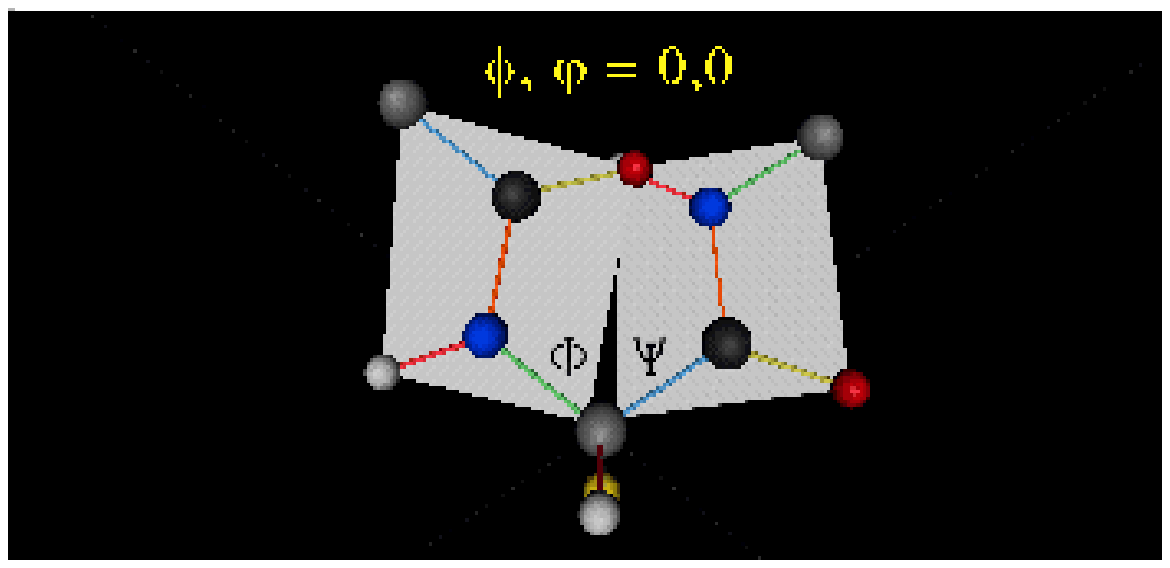}}
\hfill \caption{(Color online) Pauling and Ramachandran revisited: The top row
depicts the `classic' structures of a $\alpha$-helix (a) and a
pleated $\beta$-sheet (b). The main-chain backbone atoms and the
$C^{\beta}$ atoms of the side chain groups are shown (color codes
are different for (a) and (b)). Hydrogen bonds, which stabilize
the structures are shown as dashed lines. In the bottom row we
show the Ramachandran plot (c) describing how the torsional
degrees of freedom ($\psi$,$\phi$), the backbone dihedral angles
within an all atom representation, are constrained by steric
effects. The colored areas in the plot correspond to allowed
regions in conformational space. The structures (a) and (b)
stabilized by hydrogen bonding indeed lie squarely within the
sterically accessible regions. An example of a dipeptide
conformation disallowed because of steric hindrance is shown in
(d).} \label{Fig12}
\end{figure}

We conclude by revisiting the classic theoretical work of
Pauling\cite{Pauling1,Pauling2} and Ramachandran\cite{Rama}.  Both of
them considered the protein backbone which is the common part of all
proteins.  Pauling and his coworkers explored the types of structures
that are consistent with both the backbone geometry and the formation
of hydrogen bonds.  They predicted that helices and sheets are the
structures of choice in this regard
(Fig. \ref{Fig12}(a,b)). Ramachandran and his coworkers carried out
their pioneering work more than a decade after Pauling.  They
considered the role of excluded volume or steric interactions between
the adjacent amino acids in reducing the available conformational
phase space (Fig. \ref{Fig12}(c)).  Astonishingly, the two
significantly populated regions of the Ramachandran plot correspond to
the $\alpha$-helix and the $\beta$-strand.  Even though hydrogen bonds
and sterics are not related to each other, they are both promoters of
helices and sheets.  Is this concurrence of events a mere accident?
The marginally compact phase of short tubes has helices and sheets as
its preferred structures.  In order for Nature to take advantage of
this phase of matter, proteins, which obey physical law, may have been
selected to conform to the tube geometry.  Hydrogen bonds serve to
enforce the parallelism of nearby tube segments, a feature of both
helices and sheets while sterics emphasizes the non-zero thickness of
the tube and serves to position it in the marginally compact phase.
Because the marginally compact phase is a finite size effect, proteins
tend to be relatively short compared to conventional macromolecules
including DNA.  Indeed, proteins seem to be a vivid example of the
adaptation of Nature to her own laws.

In his insightful book, `The Fitness of the Environment',
Henderson extended the notion of Darwinian fitness to argue that
``the fitness of environment is quite as essential a component as
the fitness which arises in the process of organic evolution."
Strikingly, the chemistry of proteins ensures that they are
self-tuned to occupy the marginally compact phase of short tubes.
One cannot but marvel at how several factors, the steric
interactions; hydrogen bonds which provide the scaffolding for
protein structures; the constraints placed by quantum chemistry on
the relative lengths of the hydrogen and covalent bonds and the
near planarity of the peptide bonds; and the key role played by
water all reinforce and conspire with each other to place proteins
in this novel phase of matter.

Proteins have proved to be difficult to understand because of
their inherent complexity with twenty types of amino acids and the
role played by water, because they are relatively short molecules
compared to generic man-made polymers and are therefore likely to
be characterized by `non-universal' behavior, and because of the
complexities associated with the random process of evolution.
Nevertheless, our work suggests that there is an underlying
stunning simplicity. While sequences and functionalities of
proteins evolve, the folds that they adopted, which in turn
determine function, seem to be determined by physical law and are
not subject to Darwinian evolution.  In that regard, these folds
may be thought of as immutable or Platonic. Protein folds do not
evolve -- rather, the menu of possible folds is determined by
physical law. In that sense, it is as if evolution acts in the
theater of life and shapes sequences and functionalities but does
so within the fixed backdrop of the Platonic folds.

Henderson\cite{Henderson} wrote ``The properties of matter and the
course of cosmic evolution are now seen to be intimately related
to the structure of the living being and to its activities; they
become, therefore, far more important in biology than has been
previously suspected. For the whole evolutionary process, both
cosmic and organic, is one, and the biologist may now rightly
regard the universe in its very essence as biocentric."  His
intriguing ideas continue to provoke thought even as we strive to
understand the connections between life and the laws of nature.

\vspace{1cm}

\noindent {\bf Acknowledgements} We are indebted to Phil Anderson,
Buzz Baldwin, Paul Chaikin, Hue-Sun Chan, Marek Cieplak, Morrel
Cohen, Michael Denton, Ken Dill, Chris Dobson, Russell Doolittle,
Sam Edwards, Alessandro Flammini, Leo Kadanoff, Arthur Lesk, Tim
Lezon, Davide Marenduzzo, Cristian Micheletti, Harold Scheraga,
Michele Vendruscolo and Saraswathi Vishveshwara for useful
discussions and especially to George Rose for teaching us much
about proteins. This work was supported by PRIN 2003, FISR 2001,
INFM, NASA, NSF IGERT grant DGE-9987589, The Templeton Foundation
and NSF MRSEC at Penn State.

\appendix

\section{Three-body description of a tube}

In this appendix we will describe how a suitable three-body
potential \cite{JSP} characterizes the self-avoidance of a tube of
thickness $\Delta$ whose axis, $\mathcal{C}$, is a {\em smooth}
curve, ${\bf r}(s)$, parametrized by its arc-length $s$ with $0\le
s \le L$, $L$ being the total length of the tube. The tube is a
one dimensional generalization of the zero dimensional hard sphere
case as described in the text.  The self-avoidance of an ensemble
of hard spheres, each of radius $\Delta$, can be ensured by
requiring that none of the distances between all pairs of sphere
centers is less than $2\Delta$.

Let us consider, first, a closed curve, i.e. ${\bf r}(0) = {\bf
r}(L)$.  At each position, $s$, along the curve $\mathcal{C}$ we
consider an infinitesimally thin circular disk of radius $\Delta$,
$\Sigma(s,\Delta)$, centered at the point ${\bf r}(s)$ and
perpendicular to the tangent vector $d{\bf r}(s)/ds$ at $s$. The tube
is simply the union of all the disks.  The self avoidance is imposed
by requiring that pairs of disks at different points do not intersect,
$\Sigma(s,\Delta)\bigcap \Sigma(s',\Delta) = \emptyset \quad \forall
s, s'$.

There is an easier way to implement the self-avoidance (steric
constraints), which underscores the key difference between the
hard sphere and the tube problem. Indeed, in the latter case,
there are two classes of lengths which are relevant to the steric
interaction: the radius of curvature, $\mid\ddot{{\bf
r}}(s)\mid^{-1}$, at each position $s$ and the closest approach
distances (note that $\mid\dot{{\bf r}}(s)\mid=1$ within the
arc-length parametrization). A closest approach occurs at, say,
points ${\bf r}(s_1)$ and ${\bf r}(s_2)$ ($s_1 \ne s_2$) when
${\bf r}(s_1) - {\bf r}(s_2)$ is perpendicular to both tangent
vectors at $s_1$ and $s_2$. For a {\em smooth} closed curve there
is at least one such closest approach. It is rather intuitive
\cite{Buck,Litherland} that a necessary and sufficient condition
for the self avoidance is that $\Delta$ be less than the minimum
among $\mid\ddot{{\bf r}}(s)\mid^{-1} \: \forall s$ and $1/2
\mid{\bf r}(s_1) - {\bf r}(s_2)\mid \: \forall s_1, s_2$ where
$\dot{{\bf r}}(s_i), i=1,2$ are both perpendicular to ${\bf
r}(s_1) - {\bf r}(s_2)$. This minimum is called the thickness, $
\Delta({\mathcal{C}})$, of the curve $\mathcal{C}$
\cite{Buck,Litherland}. The minimum among the closest approach
distances is analogous to the minimum among all distances between
pair of centers in the hard sphere problem. The fact that the tube
is a linear object introduces another length in the problem which
is the minimum among all radii of curvature and it is local in
nature in the sense that it involves nearby points of the curve
$\mathcal{C}$: the radius of curvature at position $s$ represents
the radius of the circle that best approximates the curve
$\mathcal{C}$ at $s$.

We now turn to a reformulation of the tube self avoidance
constraint in a much more appealing way that makes it more similar
to the self-avoidance recipe for hard spheres. Following ref.
\cite{GM} let us consider a triplet of positions along the curve
$\mathcal{C}$ (instead of a pair of centers as in the hard sphere
problem), ${\bf r}_i \equiv {\bf r}(s_i), i=1,2,3$. These
positions define a plane and hence a unique circle through them
whose radius is \begin{equation} r({\bf r}_1,{\bf r}_2,{\bf r}_3)
= \frac{\mid{\bf r}_2-{\bf r}_1 \mid \mid{\bf r}_3-{\bf r}_1\mid
\mid{\bf r}_3-{\bf r}_2 \mid} {4 A({\bf r}_1,{\bf r}_2,{\bf r}_3)}
\end{equation} where $A({\bf r}_1,{\bf r}_2,{\bf r}_3)$ is the area
of the triangle whose vertices are ${\bf r}_1,{\bf r}_2$ and ${\bf
r}_3$. The theorem proved in reference \cite{GM} states that

\begin{equation}\label{min}
\Delta '({\mathcal{C}})\equiv \min_{s_1,s_2,s_3}r({\bf r}(s_1),{\bf
r}(s_2),{\bf r}(s_3)) = \Delta ({\mathcal{C}})
\end{equation}
where the $s$'s do not need to be distinct. Indeed it is easy to show
that when $s_1,s_2,s_3 \rightarrow s$ then $r({\bf r}(s_1),{\bf
r}(s_2), {\bf r}(s_3))\rightarrow \mid \ddot{{\bf r}(s)}\mid^{-1}$,
the radius of curvature at $s$. Furthermore it is not difficult to
show that the search for the minima in eq.(\ref{min}) can be
restricted to

\begin{equation}\label{min2}
\lim_{s_2 \rightarrow s_1} r({\bf r}(s_1),{\bf r}(s_2),{\bf r}(s_3))
\equiv r({\bf r}_1,{\bf r}_1,{\bf r}_3)
\end{equation}
which is the radius of the circle through the point ${\bf r}(s_3)$ and
${\bf r}(s_1)$ and tangent to the curve at the latter point.

Let us assume that the minimum in eq.(\ref{min}) is reached at
three distinct points ${\bf r}_1,{\bf r}_2,{\bf r}_3$ and let us
consider the sphere of radius $\Delta '({\mathcal{C}})$,
eq.(\ref{min}), through them.  If it is not tangent to the curve
in at least two of the points ${\bf r}_1,{\bf r}_2,{\bf r}_3$ we
have a contradiction. Indeed if the sphere is tangent to the curve
at one or none of the three points we can shrink the sphere
slightly still keeping three intersections with the curve. However
this is a contradiction since due to the definition of thickness,
eq.(\ref{min}), any sphere of radius less than $\Delta
'({\mathcal{C}})$ cannot intersect the curve in more than two
points. Thus say that one of the points where the tangency occurs
is ${\bf r}_1$. Since the circle through ${\bf r}_1$ and ${\bf
r}_2$ and tangent to the former lies on the sphere it implies that
$r({\bf r}_1,{\bf r}_1,{\bf r}_2) \le \Delta '({\mathcal{C}})$
which is a contradiction unless the equality holds. This
demonstrates that the minimum in eq.(\ref{min}) is never
exclusively reached at three distinct points along the curve.  The
above argument leads also to the proof of the theorem. In fact if
the other tangency point is, say, ${\bf r}_2$, then in addition to
$r({\bf r}_1,{\bf r}_1,{\bf r}_2) = \Delta '({\mathcal{C}})$, one
also has $r({\bf r}_2,{\bf r}_2,{\bf r}_1) = \Delta
'({\mathcal{C}})$. One may immediately prove that this can occur
only if the tangent vectors $\dot{{\bf r}}(s_i), i=1,2$ are
perpendicular to ${\bf r}(s_1) - {\bf r}(s_2)$. Thus
$\min_{s_1,s_2,s_3}r({\bf r}(s_1),{\bf r}(s_2),{\bf r}(s_3))$
captures simultaneously both the radius of curvature and the
distances of closest approaches, consequently proving the equality
(\ref{min}).

The local thickness of the tube (global radius of curvature in
ref. \cite{GM}), at each ${\bf r}\left(s\right)\in\mathcal{C}$,
may be defined as

\beq \Delta_{{\bf
r}\left(s_1\right)}\left(\mathcal{C}\right)=\min_{s_2,s_3}
r\left({\bf r}\left(s_1\right), {\bf r}\left(s_2\right),{\bf
r}\left(s_3\right)\right).   \label{globrad} \eeq

Of course the thickness $\Delta\left(\mathcal{C}\right)$ is the
minimum of $\Delta_{{\bf
r}\left(s\right)}\left(\mathcal{C}\right)$ as ${\bf
r}\left(s\right)$ varies on $\mathcal{C}$. Another theorem proved
in ref. \cite{GM} states that if $\mathcal{C}$ can be deformed
smoothly in order to maximize the thickness without changing the
knot type, the resulting curve, $\mathcal{C}^{*}$, called ``ideal
shape'' of the given knot type, has $\Delta_{{\bf
r}\left(s\right)}\left(\mathcal{C}^{*}\right)=
\Delta\left(\mathcal{C}\right)$ for all points where
$\mid\ddot{{\bf r}}(s)\mid\ne0$. Fig. \ref{Fig5} is a histogram of
local thicknesses for a sample of native protein structures. The
variations of the local thickness around the average value
$2.7$ \AA\/ is about $7$ \%.

What we learn from the above mathematical framework is that a mere
pairwise interaction does not suffice to describe the steric
constraint of a tube whose axis is a string $\mathcal{C}$
\cite{JSP}. This is because, in addition to the distance between
two points on a string, one also needs to know the context, i.e.
the local direction of the string in the proximity of the points
themselves. Let us consider a three-body potential $V(r({\bf
r}_1,{\bf r}_1, {\bf r}_3))$ characterizing the interaction
between three particles on the axis of the string in terms of the
radius of the circle through them (notice that this potential is
invariant under translation, rotation and permutation of the three
points). $V(r)$ could be the same as commonly used in the hard
sphere problem, i.e. $V(r) = \infty$ when $r < \Delta$ and $V(r) =
0$ otherwise (in the hard sphere problem $r$ is half of the
distance between a pair of sphere centers). This length scale
neatly solves the contextual problem mentioned above.  When two
parts of a chain come together, the radius of a circle passing
through two of the particles on one side of the chain and one
particle from the other side of the chain turns out to be a
measure of the distance of approach of the two sides of the chain.
On the other hand, when one considers three particles
consecutively along the chain, the radius of the circle passing
through them is simply the local radius of curvature.  Indeed when
three such particles form a straight line, the radius goes to
infinity and the three particles essentially become
non-interacting.  The straight line configuration is the best that
the particles can do in terms of staying away from each other
given that they are constrained to be neighbors along the chain.
In the case of a polymer chain, such as a protein, a tube whose
axis is a {\em smooth} string is clearly an approximation.  One
ought to introduce a {\em discrete} curve $\{ {\bf r}_1, {\bf
r}_1, \ldots, {\bf r}_N \}$, and the continuous variable $s$ of
the curve now becomes discrete. In correspondence with the
considerations above, one may again define the thickness of a
discrete curve, $\mathcal{C}$, as \cite{GM}

\begin{equation}
\Delta ({\mathcal{C}}) = \min_{i,j,k}r({\bf r}_i,{\bf
r}_j,{\bf r}_k)
\label{Disc}
\end{equation}
where now $i, j$ and $k$ are all distinct. For a discrete curve there
is no guarantee that the minimum is obtained from among $r({\bf
r}_i,{\bf r}_j,{\bf r}_k)$ with at least two of the three indices
separated by one unit (e.g. $j = i \pm 1$), but one can still
distinguish between a local and a non-local contribution to the
thickness, according to whether $i,j,k$ are consecutive along the
chain or not. In the latter case, the minimum obtained from among
$r({\bf r}_i,{\bf r}_j,{\bf r}_k)$ gives half the minimum distance of
closest approach computed for the discrete chain. Similarly, there is
no simple restriction of the triplets when one deals with open
continuous curves with free ends.

\section{Tube versus string and beads model}

In this appendix we summarize the main differences between the thick
polymer model (TP) that we deal with in this work and the Edwards'
model \cite{DOI} (EM) in the presence of a bending rigidity term (the
analogue of the Edwards' model in the discrete case is the usual
string and beads model). In both cases, one may add a twist rigidity
term, which we neglect here, for simplicity. Let us consider the case
of continuous chains.  The Hamiltonian for the generalized EM model is
\beqa {\cal H}_{\rm EM}(\{{\bf r}\}) & = & \frac{1}{2} \int_0^L
\dot{{\bf r}}(s)^2 ds + \frac{\kappa_b}{2} \int_0^L \ddot{{\bf
r}}(s)^2 ds + \nonumber \\ & + & \frac{v_2}{6} \int_0^L\int_0^L
\delta({\bf r}(s)-{\bf r}(s'))ds\, ds' + \nonumber \\
& + & \; \frac{v_3}{90} \int_0^L\int_0^L\int_0^L
\delta({\bf r}(s)-{\bf r}(s')) \cdot \nonumber \\
& & \qquad \quad \cdot \delta({\bf r}(s)-{\bf r}(s'')) ds\, ds'\, ds''\, .
\label{EM}
\eeqa
The self-avoidance in the TP model\cite{note13} is given by

\beqa {\cal H}_{\rm TP}(\{{\bf r}\}) & = &
\int_0^L\int_0^L\int_0^L V\left(R_c\left({\bf r}(s),{\bf
r}(s'),{\bf r}(s'')\right)\right) \cdot \nonumber \\ & & \qquad \qquad
\qquad \qquad \qquad \cdot \; ds\, ds'\, ds'' \, , \label{TP} \eeqa
where $R_c\left({\bf r}(s),{\bf r}(s'),{\bf r}(s'')\right)$ is the
radius of the circle  through the three points ${\bf r}(s)$, ${\bf
r}(s')$, ${\bf r}(s'')$ and \beq\label{3bodycontchain}
V\left(r\right)= \left \{
\begin{array}{cc}
 \infty & \mbox{ if $r < R_0$ } \\
 -1   & \mbox{ if $ R_0 < r <  R_1$ } \\
  0   & \mbox{ if $ R_1 < r  $ }
\end{array}
\right.
\eeq

Note that in the limit of a continuous chain the EM model needs the
introduction of singular potentials, in order to deal with the fact that a
two-body potential is unable to distinguish whether two nearby
beads are far apart or not along the chain. Within the context of the
Edwards' model such singularities can then be treated successfully
within a perturbative renormalization group
approach\cite{CLOISEAUX}. On the other hand, the need of singular
potentials is deftly avoided when using the three-body prescription
implied by the thickness constraint in the TP model\cite{note13}.

The details of the discretization scheme matter for a discrete
chain.  First, the discretization introduces a natural cut-off
length scale. Second, the three-body potential $V$
of Equation \ref{TP} cannot be used by itself
for a discrete chain.  Indeed, in the absence of
a two-body repulsion, the
chain would collapse onto a circle of radius between $R_0$
and $R_1$ and would wind repeatedly along it.

{\bf High temperature phase} It is well known that in the high
temperature regime the critical behavior of the EM in the limit of
very long chains is governed by the exponent $\nu\simeq0.58$, so that
a typical length $\xi$ measuring the spatial extension of the chain
scales as $\xi\sim L^{\nu}$, where $L$ is the chain length. The chain
is {\em swollen} with respect to the Gaussian random walk behavior for
which $\nu=1/2$. The same feature holds for the TP; in the high
temperature regime the different symmetry properties induced by the
inherent anisotropy of a thick tube are averaged out, and a chain of
coins shares the same critical behavior as a chain of spheres.

Interestingly, other features such as the form of the two-point
tangent-tangent correlation function along the chain differentiate
the TP from the EM. In the absence of twisting rigidity (the {\em
intrinsic} twist of the chain, as defined by the {\em torsion} of
the corresponding curve for the EM or the axis of the tube for the
TP, is described by the energy term $\frac{\kappa_t}{2} \int_0^L
\dot{\hat{b}}(s)^2 ds$, where $\hat{b}(s)$ is the {\em binormal}
vector which is part of the Frenet triad), one gets a simple
exponential decay in both cases.  However when the twisting
rigidity $\kappa_t$ is introduced, the EM exhibits an oscillatory
decaying correlation function for {\em any} value of $\kappa_t$,
whereas the TP  crosses over from simple to oscillatory decay on
increasing $\kappa_t$ (see Appendix F) \cite{Uslast}. The
existence of a transition line in the parameter space
$\left(\kappa_b,\kappa_t\right)$ separating simple from
oscillatory decay is a novel feature of the TP model.

{\bf Persistence length} Another similarity between the TP and the EM
is the following: at any {\em fixed} value of the length $L$ of the
chain and of the temperature $T$, the persistence length $l_p$ (which
is a measure of the distance along the chain after which the tangent
vectors become uncorrelated) diverges both for the TP, in the limit
$\Delta\rightarrow\infty$ of infinite thickness, and for the EM, in
the limit $\kappa_b\rightarrow\infty$ of infinite bending rigidity.
The thickness constraint indeed stiffens the chain locally.  Yet, a
closer look reveals an important difference between the TP and the EM
model. In an ideal case in which non-local interactions are
disregarded we get a different scaling behavior. For the EM,
$l_p\sim\kappa_b /k_BT$, whereas in the TP the persistence length does
{\em not} increase at low temperatures, a first hint that the low
temperature behavior of the TP may be radically different than the EM.

\begin{figure}
\centering
\epsfig{height=5.0cm,file=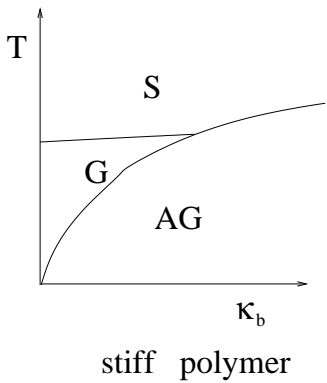}
\caption{Schematic phase diagram of a EM model in the temperature
($T$), bending rigidity ($\kappa_b$) plane \cite{STF}. The different phases are
S=Swollen, G=globule, and AG=Asymmetric globule.}
\label{FigA1}
\end{figure}

\begin{figure}
\centering
\epsfig{height=6.0cm,angle=-90,file=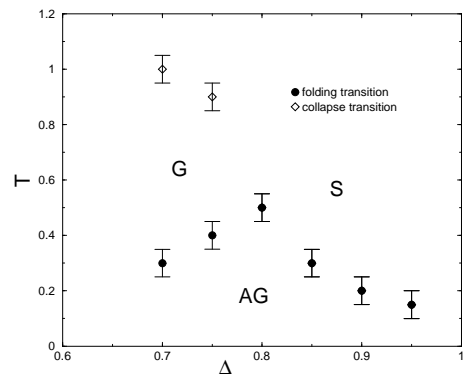}
\caption{Phase diagram for a thick polymer chain in the temperature-thickness
$(T,\Delta)$ plane obtained with Monte-Carlo simulations (see
\cite{Macro} for details). The different phases are
S=Swollen, G=globule, and AG=Asymmetric globule.}
\label{FigA2}
\end{figure}

{\bf Low temperature phase}
The anisotropy inherent in the thick tube description strongly affects
the behavior of the TP model at low temperatures, as can be seen by
comparing the bending rigidity/temperature phase diagram (Figure
\ref{FigA1}) for the EM and the corresponding thickness/temperature
phase diagram (Figure \ref{FigA2}) for the TP {\em in the
thermodynamic limit}.

Let us first note that whereas the swollen (S) and the (disordered
compact) globule (G) phase share similar features in the two cases
(but see the above discussion concerning the correlation function
properties in the swollen phase), the asymmetric globule (AG)
(semi-crystalline) phase is different. The persistence length
diverges (strictly at $T=0$) with the chain length for both EM and
TP (the chain is locally straight). For the latter, this arises
from the interplay of the thickness constraint and the interaction
promoting compaction, so that the resulting ground state
conformation will have tube segments aligned with respect to one
another similar to the Abrikosov flux lattice, filling the space
with hexagonal symmetry. For the former, this is a mere
consequence of the local bending rigidity, so that ground state
conformations will likely consist of planes stacked onto each
other with parallel (or antiparallel) alignment within the same
plane, but not necessarily between different planes.

A second crucial difference is that {\em in the limit of zero
temperature} the EM is in the AG phase for all finite values of
the bending rigidity, whereas the TP exhibits a transition from
the AG phase to the swollen phase with increasing thickness. This
has profound consequences, especially when finite size effects are
taken into account. It is instructive to revisit the phase diagram
at $T=0$ for a TP in the plane $\left(L/R,\Delta/R\right)$ (see
Fig. \ref{Fig2}). If $\Delta>R$, the chain cannot avail of the
attraction, the length of the chain does not play any role and in
the $L\rightarrow\infty$ limit one gets the critical behavior of
the swollen phase. When $\Delta<R$, in the thermodynamic limit
$L\rightarrow\infty$, the chain is in the asymmetric globule phase
resembling the Abrikosov flux lattice with hexagonal symmetry, but
novel phenomena occur for finite chain length. If $L<2\pi R$ {\em
all} parts of the chain are able to interact with each other. When
$L>2\pi R$, this is still true for small enough thickness.
However, as the thickness increases, this is not possible anymore
and the chain adopts a conformation which optimizes the attractive
interaction. In the long chain limit the boundary line between the
two regimes scales as $L/R\sim\left(R/\Delta\right)^2$.  This
result is obtained by equating the volume occupied by the tube
$L\Delta^2$ to the volume of the sphere of attraction $R^3$. For
shorter chains the compact regime at intermediate thickness in
which the chain seeks to compact itself within the constraint
of the thickness is indeed {\em marginal}, being sandwiched
between the featureless compact\cite{lastlast} and the swollen
regime described above. It is precisely in this window of
parameter space that we find marginally compact ground state
structures such as space-filling helices. This finite-size feature
of the TP model is quite robust independent of the details
introduced, for instance in the discrete case. None of these
features are present for the EM case for which there is no
dependence whatsoever on the bending rigidity at $T=0$.

\section{Optimal helix}

In this Appendix we derive the value $c^*$ of the
pitch/radius ratio $c$ of an optimal space-filling helix.
The radius of curvature of such a helix equals the tube
radius and is equal to half the minimum distance of closest
approach between different turns of the helix.

The parametric equation of a helix is
\begin{equation}
{\bf x}\left(t\right) = \left(r\cos t,r\sin t,vt\right);
\label{appeh1}
\end{equation}
where the pitch/radius ratio is $c=\frac{2\pi v}{r}$. The tangent and the
acceleration vectors are:
\begin{eqnarray}
\dot{{\bf x}}\left(t\right) & = & \left(-r\sin t,r\cos t,v\right)\:;\\
\ddot{{\bf x}}\left(t\right) & = & \left(-r\cos t,-r\sin t,0\right)\:.
\label{appeh2}
\end{eqnarray}
Since $\dot{{\bf x}}\left(t\right) \cdot \ddot{{\bf x}}\left(t\right) =
0$, the radius of curvature is simply given by
\begin{equation}
\rho_L = \frac{\left|\dot{{\bf x}}\left(t\right)\right|^2}
{\left|\ddot{{\bf x}}\left(t\right)\right|} = r\left(1 + \frac{v^2}{r^2}\right)
\label{appeh3}
\end{equation}
independently of $t$.

We define the non-local radius of curvature as half the distance of
closest approach between successive turns of the helix. Fix a point
${\bf A} = {\bf x}\left(t\right)$ on the curve, and compute the distance
$d\left(s,t\right) = \left|{\bf B} - {\bf A}\right|$ from a second point
${\bf B} = {\bf x}\left(s\right)$ moving along the curve as a function
of $s$. The non-local radius is then
\begin{equation}
\rho_{NL}\left(t\right) = \frac{1}{2} \min_{s\ne t}\left\{d\left(s,t\right)\right\}\; ,
\label{appeh4}
\end{equation}
with the requirement that $\frac{\partial d\left(s,t\right)}{\partial
s}=0$ at some $s^*\ne t$, implying that ${\bf B} - {\bf A}$ is
perpendicular to the tangent vector $\dot{{\bf x}}\left(s^{*}\right)$. Note
that the non-local radius need not exist (for open curves) and is in
principle a varying function of $t$, when the curve is not invariant
under translation along it.

Because the helix is invariant under translation along the curve,
implying that ${\bf B} - {\bf A}$ is perpendicular also to the
tangent vector $\dot{{\bf x}}\left(t\right)$, we can choose $t=0$
so that
\begin{equation}
d^2\left(s\right) \equiv d^2\left(s,0\right) =
r^2\left[2\left(1-\cos s\right) + \frac{v^2}{r^2} s^2 \right]\; .
\label{appeh5}
\end{equation}
The condition allowing to get extremal points for $d^2\left(s\right)$ is
\begin{equation}
\sin s + s \frac{v^2}{r^2} = 0\ .
\label{appeh6}
\end{equation}
One trivial solution of this equation is $s=0$ and there is no
other solution for sufficiently high pitch to radius ratio
$c=\frac{2\pi v}{r}$. If $c$ is decreased, new solutions appear,
two at a time, the smaller a maximum and the greater a minimum,
corresponding to the increasing packing of helix turns. We are
interested in the minimum $s^*$ corresponding to ${\bf A}$ and
${\bf B}$ staying on two consecutive turns, that is $\pi < s^* <
2\pi$. For sufficiently low $c$, the above equation then defines
the implicit function $s^*\left(c\right)$, and one has
\begin{equation}
\rho_{NL} = \frac{1}{2} d\left[s^*\left(c\right)\right]
\label{appeh7}
\end{equation}
for all points of the helix. In the limit $c\ll 1$, one has
$s^*\simeq2\pi$ and thus $\rho_{NL} \simeq \pi v = \frac{p}{2}$, where
$p = 2\pi v$ is the pitch of the helix, as expected. The particular
value $c=c^*$, for which the local and the non-local radius of
curvature are equal, is then defined by
\begin{equation}
\frac{d\left[s^*\left(c^*\right)\right]}{2r} = 1 + \left(c^*\right)^2\; .
\label{appeh8}
\end{equation}

Thus, according to the definition of thickness given in Appendix A,
$\Delta_{\rm helix}=\rho_L$ if $c>c^*$, since the radius of curvature
is smaller than the non-local radius. A tube swelling around the helix
would stop increasing due to local singularities, leaving space
between the successive turns of the helix. On the other hand, the
non-local radius is smaller than the radius of curvature if $c<c^*$,
implying $\Delta_{\rm helix}=d\left(s^{*}\left(c\right)\right)/2$. In
such a case, the tube would stop swelling due to self-intersection
between different turns, leaving a hole in the middle of the helix. At
$c=c^*$, one obtains an optimal space-filling helix with a special
pitch to radius ratio of $c^* \approx 2.512$ (shown in
Fig. \ref{Fig4}).

\section{Geometrical constraints determined from experimental data}

\begin{figure}
\centering
\hfill
\subfigure[]
{\includegraphics[width=3.3cm,angle=-90]{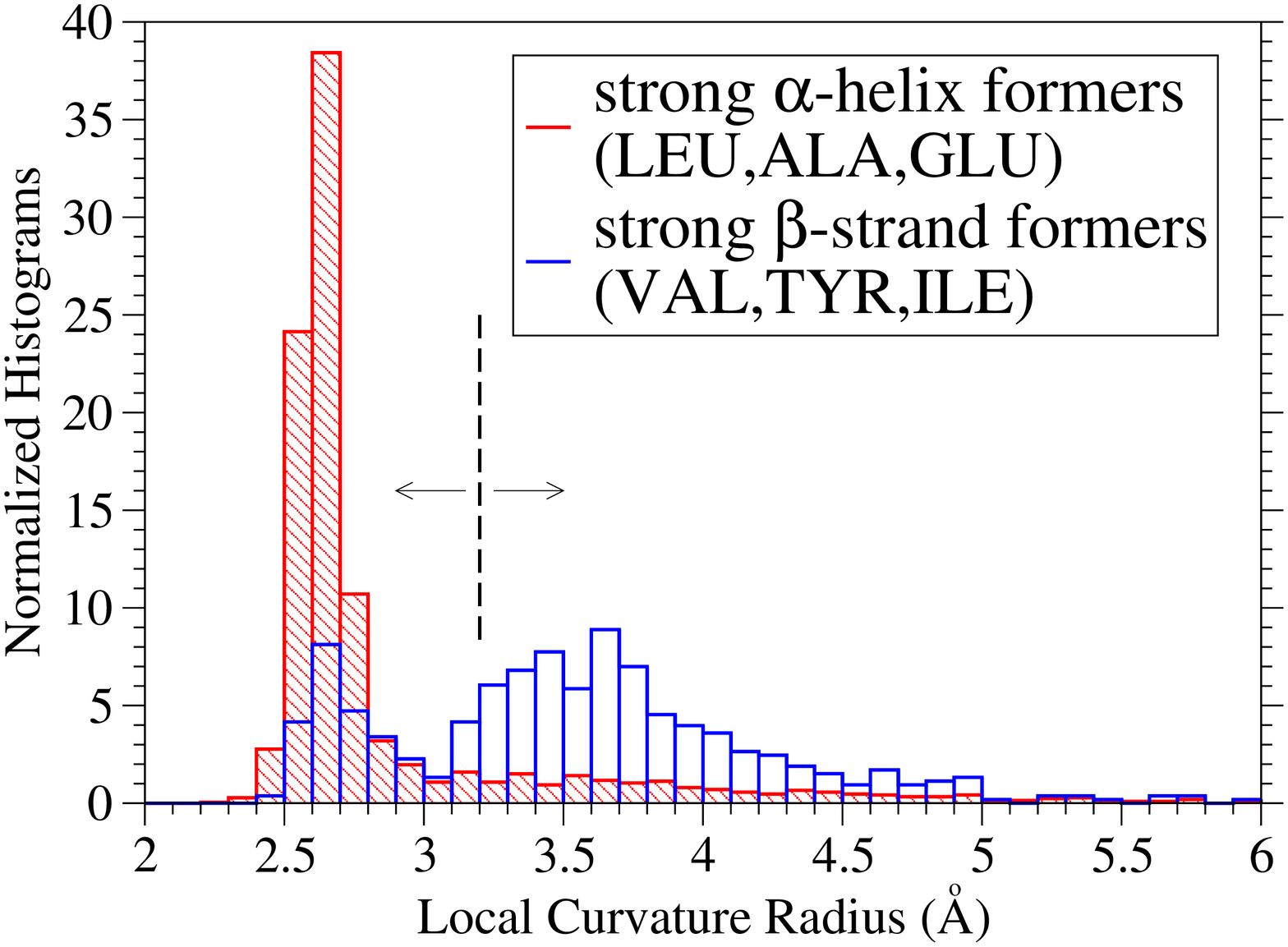}}
\hfill
\subfigure[]
{\includegraphics[width=3.3cm,angle=-90]{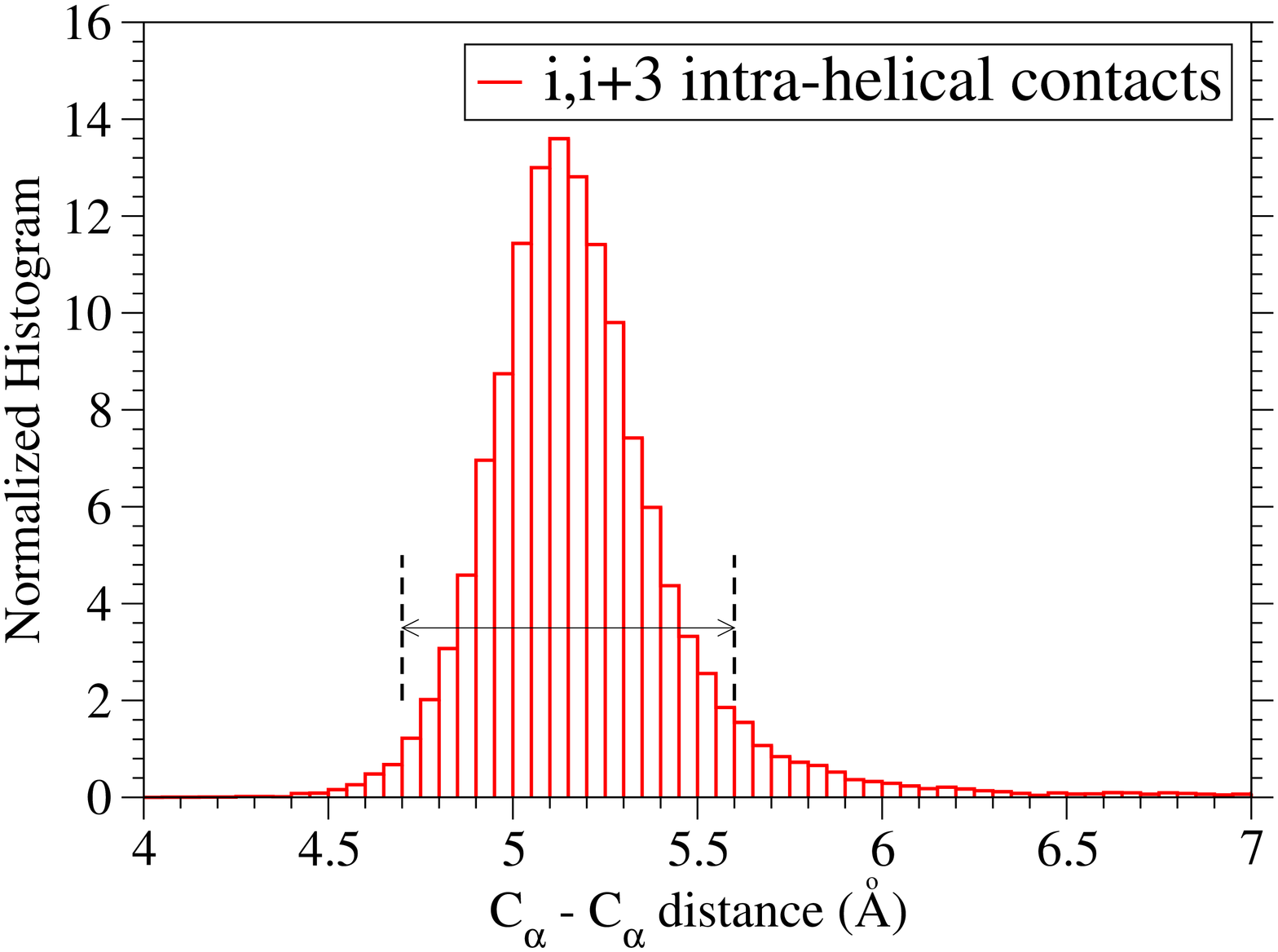}}
\hfill
\\
\hfill
\subfigure[]
{\includegraphics[width=3.3cm,angle=-90]{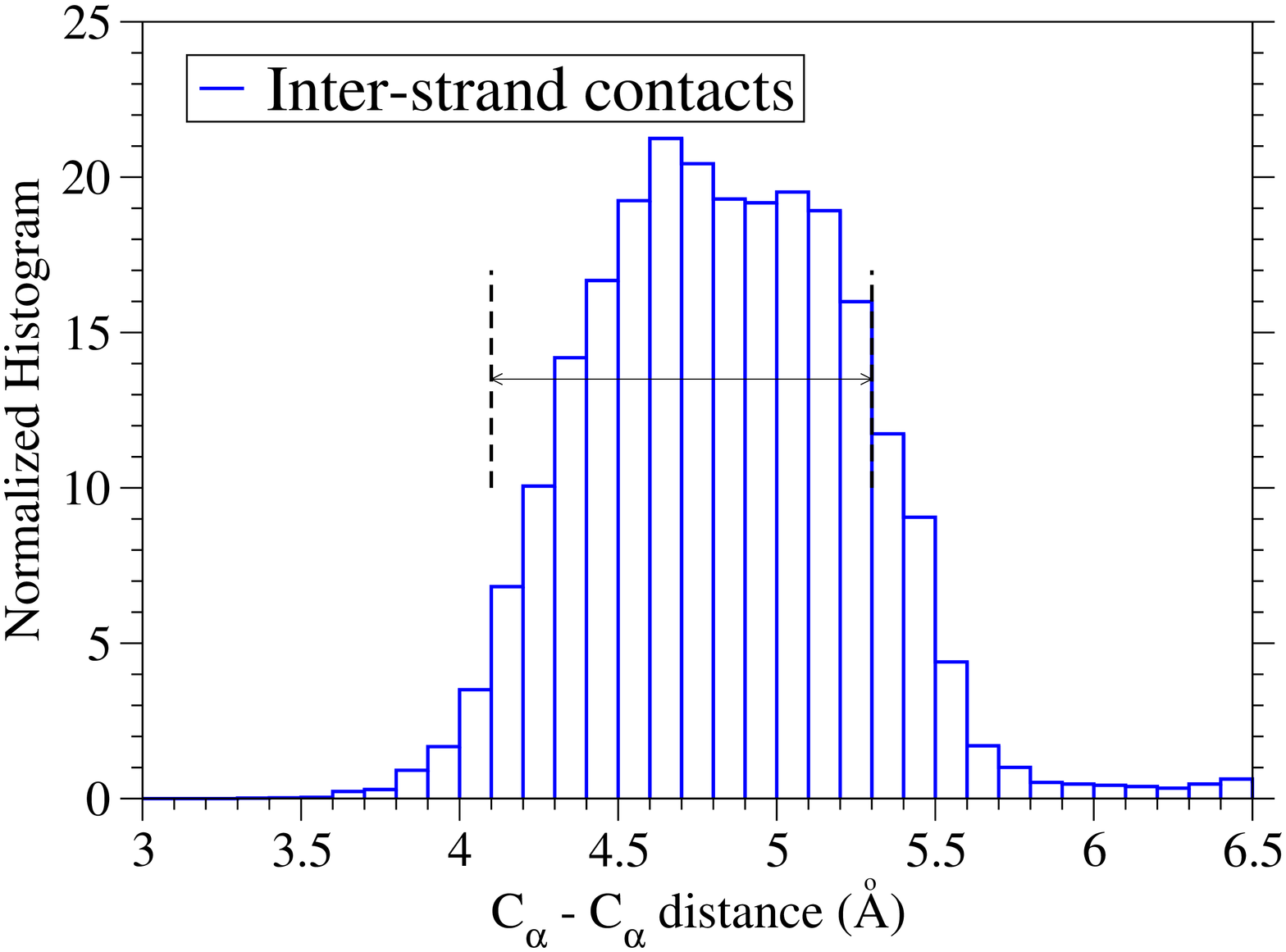}}
\hfill
\subfigure[]
{\includegraphics[width=3.3cm,angle=-90]{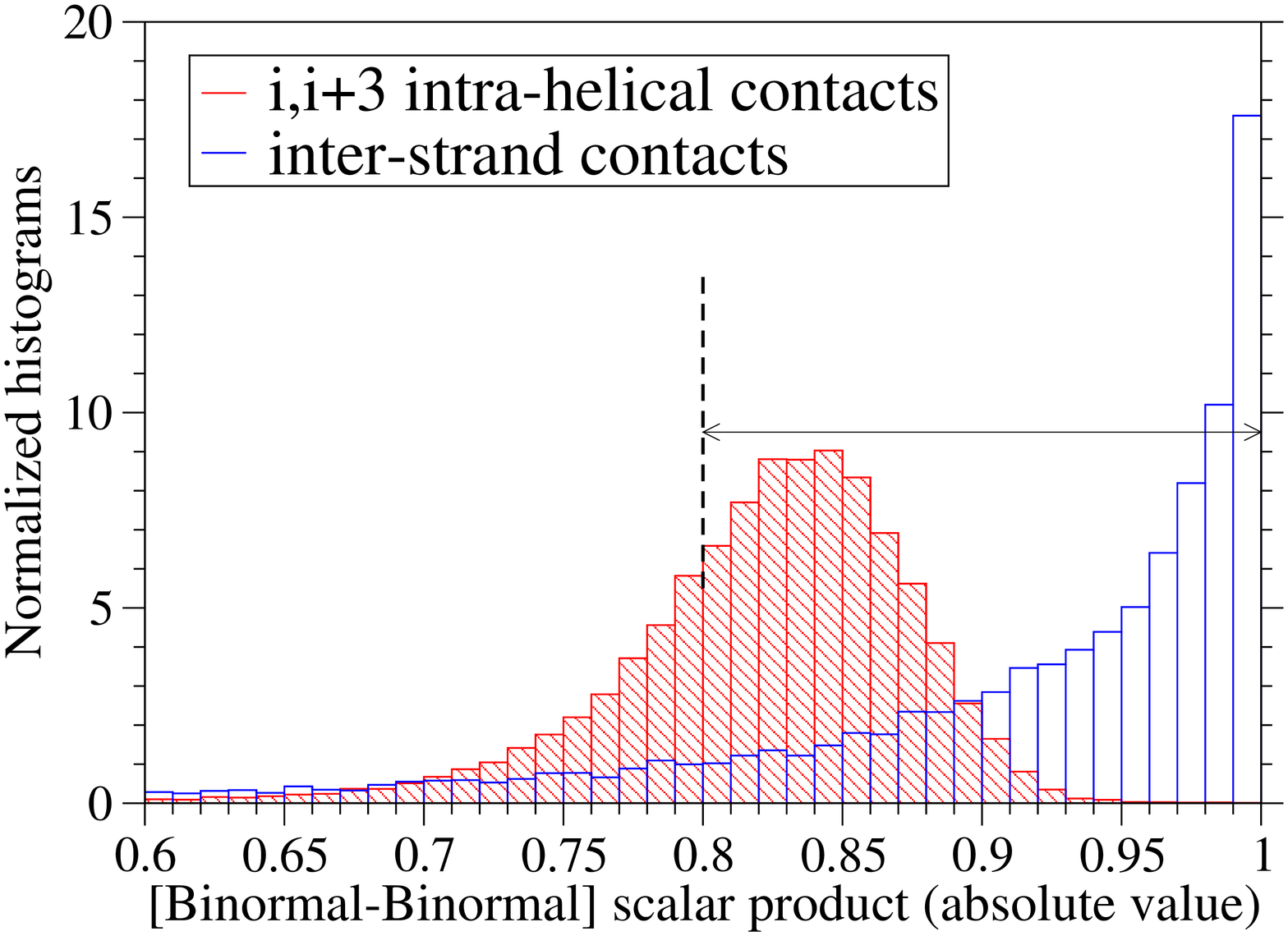}}
\hfill
\\
\hfill
\subfigure[]
{\includegraphics[width=3.3cm,angle=-90]{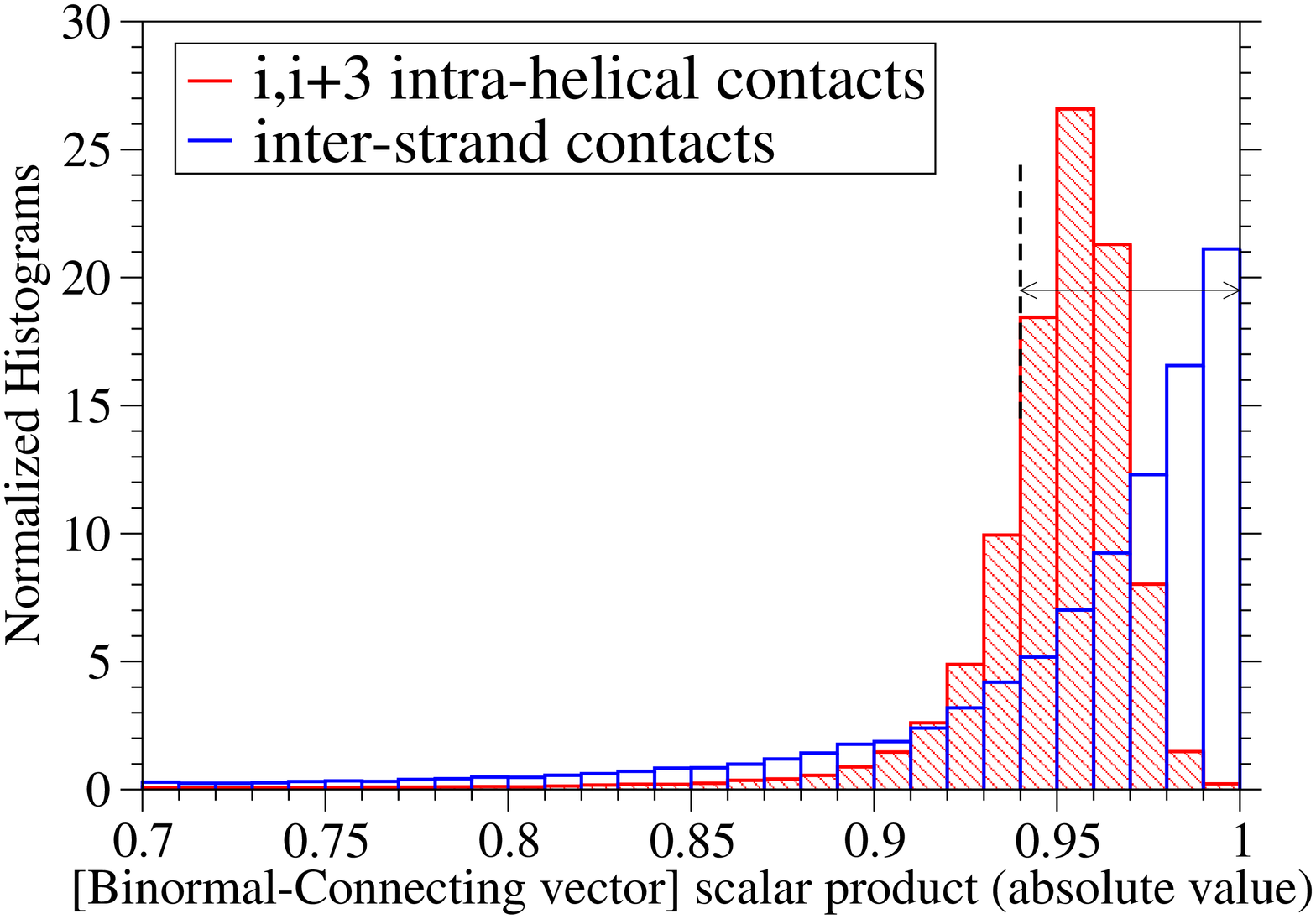}}
\hfill
\subfigure[]
{\includegraphics[width=3.3cm,angle=-90]{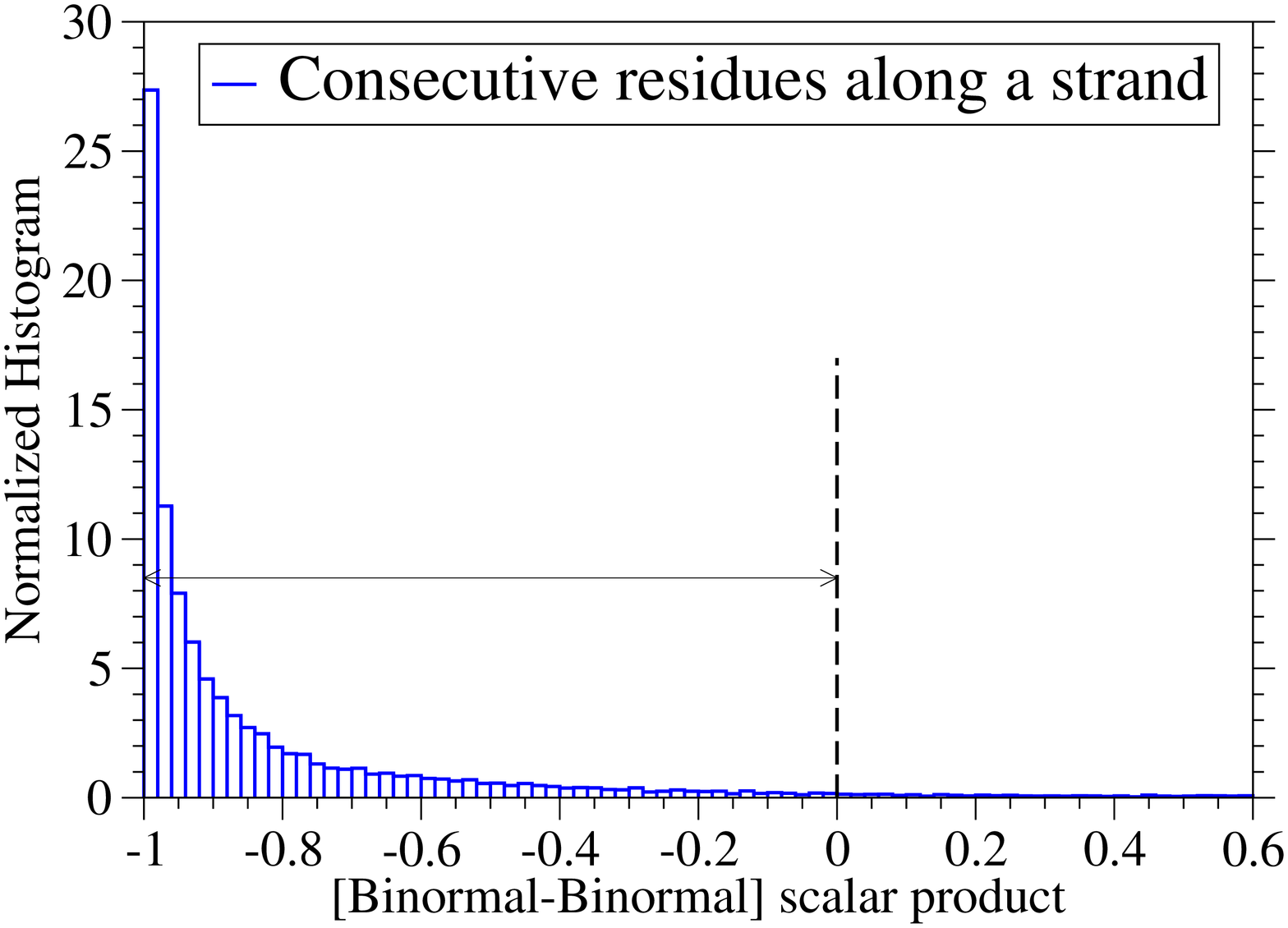}}
\hfill
\caption{(Color online) Statistical analysis of several quantities computed for
residues classified as participating in secondary structures in
protein native state structures from the Protein Data Bank. Red (blue)
histograms refer to residues participating in $\alpha$-helices
($\beta$ strand).}
\label{FigA3}
\end{figure}

In this Appendix we describe the data analysis used to elucidate the
geometrical constraints imposed by sterics and hydrogen bonds (see
Fig. \ref{FigA3}). We have used a database of 600 different protein
native structures\cite{Iksoo} consisting of sequences varying in
length from 44 to 1017, with low sequence homology and covering many
different three-dimensional folds according to the Structural
Classification of Proteins (SCOP) scheme\cite{Murzin}.  Panel (a)
depicts the histogram of the local radius of curvature associated with
two classes of triplets, the first (shown in red) featuring strong
$\alpha$-helix forming amino acids (LEU, ALA, GLU) and the second
(shown in blue) featuring $\beta$-strand formers (VAL, ILE,
TYR)\cite{Creighton} and underscores the vital role of chemistry in
choosing from among the menu of native state folds. The vertical
dashed line indicates the threshold length scale chosen in the model
for the curvature energy penalty. The remaining panels show histograms
for several quantities involved in the definition of hydrogen bonds:
the $C^{\alpha}$-$C^{\alpha}$ distance between $i,i+3$ atoms given
that $i$, $i+1$, $i+2$, $i+3$ all belong to a helix (Panel (b)) and
between $i,j$ (with $j>i+4$) atoms given that $i$, $j$, belong to a
$\beta$-strand (Panel (c)); the scalar products $\hat b_i \cdot \hat
b_j$ (Panel (d)) and $(\hat b_i + \hat b_j)\cdot \hat r_{ij}/2$
(Panel (e)) for $i,j$ contacts (with $|j-i| = 3$ (red) and with $|j-i|
> 4$ (blue) provided that no closer interstrand contact is present
among $i\pm1,j\pm1$) ($\hat b_i$ is the binormal vector at atom $i$
and $\hat r_{ij}$ is the vector joining atoms $i$ and $j$ normalized
to unit length); and the scalar product $\hat b_i \cdot \hat b_{i+1}$
for consecutive residues along a $\beta$-strand (Panel (f)). In each
case, the dashed lines and arrows depict the approximate constraints
used in our model.  All histograms are normalized in such a way that a
flat distribution would have a constant unit height.

\section{Details of model and Monte-Carlo simulations}

The protein backbone is modeled as a chain of
$C^{\alpha}$ atoms with a fixed distance of $3.8$ \AA\/ between
successive atoms along the chain, an excellent assumption for all but
non-cis Proline amino acids\cite{Creighton}.
The geometry imposed by
chemistry dictates that the bond angle associated with three
consecutive $C^{\alpha}$ atoms is between $82^{\circ}$ and
$148^{\circ}$.

\noindent
{\bf Tube geometry.}
Self-avoiding conformations of the tube whose axis is the protein
backbone are identified by considering all triplets of $C^{\alpha}$
atoms and drawing circles through them and ensuring that none of their
radii is smaller than the tube radius\cite{JSP}.  At the
local level, the three body constraint ensures that a flexible tube
cannot have a radius of curvature any smaller than the tube thickness
in order to prevent sharp corners whereas, at the non-local level, it
does not permit any self-intersections.
The backbone of $C^{\alpha}$ atoms is treated as a flexible tube of
radius $2.5$ \AA, a constraint imposed on all (local and non-local)
three body-radii, an assumption validated for protein native
structures\cite{BMMTO2}.

\noindent
{\bf Sterics.}  Steric constraints require that no two non-adjacent
$C^{\alpha}$ atoms are allowed to be at a distance closer than $4$
\AA.  Ramachandran and Sasisekharan\cite{Rama} showed that steric
considerations based on a hard sphere model lead to clustering of the
backbone dihedral angles in two distinct $\alpha$ and $\beta$ regions
for non-glycyl and non-prolyl residues. The two backbone geometries
that allow for systematic and extensive hydrogen
bonding\cite{Pauling1,Pauling2,EisenbergPNAS} are the $\alpha$-helix
and the $\beta$-sheet obtained by a repetition of the backbone
dihedral angles from the two regions respectively\cite{BaldwinRose}.
Short chains rich in alanine residues, which are a good approximation
to a stretch of the backbone, can adopt a helical conformation in
water (see \cite{ALA} for a detailed discussion of experimental
conditions necessary to achieve this). However, when one has more
heterogeneous side chains, the helix backbone could sterically clash
with some side-chain conformers resulting in a loss of conformational
entropy\cite{Cream}. When the price in side-chain entropy is too
large, an extended backbone conformation results pushing the segment
towards a $\beta$-strand structure\cite{BaldwinRose}. These steric
constraints are approximately imposed through an energy penalty
(denoted by $e_R$) when the local radius of curvature is between $2.5$
\AA\/ and $3.2$ \AA. (The magnitude of the penalty does not depend on
the specific value of the radius of curvature provided it is between
these values.)  There is no cost when the local radius exceeds $3.2$
\AA.  Note that the tube constraint does not permit any local radius
of curvature to take on a value less than the tube radius, $2.5$ \AA.

\noindent
{\bf Hydrogen bonds.} We do not allow more than two
hydrogen bonds to form at a given $C^{\alpha}$ location.
In our representation of the protein backbone,
local hydrogen bonds form between $C^{\alpha}$ atoms separated by
three along the sequence  with an energy defined to be $-1$ unit,
whereas non-local hydrogen bonds are those that form between
$C^{\alpha}$ atoms separated by more than $4$ along the sequence
with an energy of $-0.7$.
This energy difference is based on
experimental findings that the local bonds provide more
stability to a protein than do the non-local hydrogen
bonds\cite{Sosnick}. Cooperativity effects\cite{Fain} are taken into
account by adding an energy of $-0.3$ units when consecutive hydrogen bonds
along the sequence are formed.
There is some latitude in the choice of the values of these energy parameters.
The results that we present are robust to changes (at least of the
order of 20\%) in these parameters.

\noindent
{\bf Geometrical constraints due to hydrogen bonding}.  For hydrogen
bond formation between atom $i$ and $j$, the distance between these
atoms ought to be between $4.7$ \AA\/ and $5.6$ \AA\/ ($4.1$ \AA\/ and $5.3$
\AA) for the local (non-local) case (see Fig. \ref{FigA3}(b) for the
local case).  A study of protein native state structures reveals an
overall nearly parallel alignment of the axes defined by three
vectors: the binormal vectors at $i$ and $j$ and the vector ${\bf
r}_{ij}$ joining the $i$ and $j$ atoms.  A hydrogen bond is allowed to
form only when the binormal axes are constrained to be within
$37^{\circ}$ of each other, whereas the angle between the binormal
axes and that defined by ${\bf r}_{ij}$ ought to be less than
$20^{\circ}$ (see Fig. \ref{FigA3}(c)).  Additionally, for the
cooperative formation of non-local hydrogen bonds, one requires that
the corresponding binormal vectors of successive $C^{\alpha}$ atoms
make an angle greater than $90^{\circ}$ (see Fig. \ref{FigA3}(d)).
The first and the last residues of the chain are special cases since
their binormal vectors are not defined. In order for such residues to
form a hydrogen bond (with each other or with other internal residues
in the chain), it is required that the angle between the associated
ending peptide link and the connecting vector to the other residue
participating in the hydrogen bond is between $70^{\circ}$ and
$110^{\circ}$. As in real protein structures, when helices are formed,
they are constrained to be right-handed.  This is enforced by
requiring that the backbone chirality associated with each local
hydrogen bond is positive. The chirality is defined as the sign of the
scalar product $({{\bf r}}_{i,i+1} \times {{\bf r}}_{i+1,i+2}) \cdot
{{\bf r}}_{i+2,i+3}$\cite{Marek}.

\noindent
{\bf Hydrophobic interactions}. The hydrophobic (hydrophilic) effects
mediated by the water are captured through a relatively weak
interaction, $e_W$, (either attractive or repulsive) between
$C^{\alpha}$ atoms which are within $7.5$ \AA\/ of each other.
Note that
hydrogen bonds can easily be formed between the amino acid residues in
an extended conformation and the water molecules.
Within our model, the
intrachain hydrogen bond interaction introduces an effective
attraction, because water molecules are not explicitly present. The
hydrophobicity scale is thus renormalized (e.g. even when
$e_W$ is weakly positive, there could be an effective
attraction resulting in structured conformations such as a single helix or
a planar sheet). A
negative $e_W$ is, in any case, crucial for promoting the assembly of
secondary motifs in native tertiary arrangements.

Monte Carlo simulations are carried out with pivot and crankshaft moves
commonly used in stochastic chain dynamics \cite{Sokal}. A Metropolis
procedure is employed with a thermal weight $\exp\left(-E/T\right)$,
where $E$ is the energy of the conformation and $T$ is the effective
temperature.

\section{Correlation functions in the denatured state}

We will consider a polypeptide chain in a {\em phase} where the local
interactions dominate the behavior of the correlation functions, to be
studied below, at least at short and intermediate distances along the
chain. We thus neglect the steric interactions apart from the effect
that they have on neighboring nodes of the chain. The correlation
functions we will consider involve the unit vectors $\hat{t}_i $
parallel to ${\bf r}_{i+1} - {\bf r}_{i}$ and the binormal
${\hat{b}}_i \equiv (\hat{t}_i \times \hat{t}_{i-1})/\mid \hat{t}_i
\times \hat{t}_{i-1}\mid $. Note that, in order to facilitate the
calculations, our definition for the tangent vector is different from
the one used in Fig. \ref{Fig6} in Section IV. The geometrical
constraints of hydrogen bond formation are associated with the
binormal vector, whose definition is unchanged -- the binormal vector
is perpendicular to the plane defined by ${\bf r}_{i-1},{\bf r},{\bf
r}_{i+1}$. Let $\theta_{i} \in (0,\pi)$ be the angle between
$\hat{t}_i$ and $\hat{t}_{i-1}$ and $\phi_{i} \in (-\pi, \pi)$ the
angle by which the plane defined by ${\bf r}_{i-1},{\bf r}_{i}$ and
${\bf r}_{i+1}$ is rotated along the axis $\hat{t}_i $ with respect to
the plane defined by ${\bf r}_{i-2},{{\bf r}}_{i-1}$ and ${{{\bf
r}}}_{i}$. Quite generally the joint probability distribution of
angles, ${\mathcal{P}}(\theta_2, \phi_3, \theta_3, \phi_4,\dots )$
will depend on the entire ensemble of interactions including the
steric interactions. However in the {\em phase} we wish to study we
will assume that this probability distribution can be factorized, i.e.
we will consider the case where we have probability distributions
$\rho(\theta_i, \phi_i)$ for each pair of angles $\theta_i, \phi_i$
with $i = 3,4, \dots$ and

\begin{equation}
{\mathcal{P}}(\theta_2, \phi_3,\theta_3,\phi_4,\theta_4, \dots ) =
\rho_2(\theta_2)\prod_{i\ge3}\rho_i(\theta_i, \phi_i) \; ,
\end{equation}

where the contribution for the angle $\theta_2$ between the first
two vectors of the chain, $\hat{t}_1$ and $\hat{t}_2 $, has been
selected out. The average with respect to $\mathcal{P}$ will be
written as $\langle \cdot \rangle_{\mathcal{P}}$ whereas the
average with respect to $\rho_i(\theta, \phi)$ will be denoted
simply as $\langle \cdot \rangle_i$. In the case of a protein
sequence the $\rho_i(\theta_i, \phi_i)$ depends explicitly on the
type of amino-acids in the neighborhood of the $i$-th position. It
is this dependence that ultimately will determine the propensity
of a given segment of the protein sequence to be in a given
secondary structure. One can straightforwardly derive the
following recursion relations:

\begin{eqnarray}
\hat{t}_i =
& - & \hat{t}_{i-2}\frac{\sin \theta_i \cos \phi_i}{\sin \theta_{i-1}}
\nonumber \\
& + & \hat{t}_{i-1}\left(\cos \theta_i \sin \theta_i
\cos \phi_i \cot\theta_{i-1}\right)
\nonumber \\
& + & \hat{b}_{i-1} \sin \theta_i \sin \phi_i \; ,\\
\hat{b}_i = & + & \hat{t}_{i-2} \frac{\sin \phi_i}{\sin \theta_{i-1}}
- \hat{t}_{i-1} \cot\theta_{i-1} \sin \phi_i
\nonumber \\
& + & \hat{b}_{i-1} \cos \phi_i
\; .
\end{eqnarray}

If one wishes to calculate the correlation function $\left\langle{\bf
x}\cdot\hat{t}_i\right\rangle_{\mathcal{P}}$, where ${\bf x}$ is
$\hat{t}_2$, $\cot\theta_2\hat{t}_2$, $\hat{b}_2$ or any other
combination of them, then one needs to introduce other correlation
functions in order to have a closed ensemble of recursion
equations. By defining the vector

\begin{equation}\label{vec}
{\bf V}_i = \left(
\begin{array}{c}
\langle {\bf x}\cdot \hat{t}_i\rangle_{\mathcal{P}} \\
\langle {\bf x}\cdot \hat{t}_{i-1}\rangle_{\mathcal{P}} \\
\langle {\bf x}\cdot \hat{t}_i\cot\theta_i\rangle_{\mathcal{P}} \\
\langle {\bf x}\cdot \hat{b}_i\rangle_{\mathcal{P}}) \\
\end{array} \right) \; ,
\end{equation}

the recursions equations can be written in a compact form in terms
of the vectors ${\bf V}$'s and the {\em transfer} matrix
$\mathcal{T}_i$

\begin{equation}\label{rec}
{\bf V}_i = {\mathcal{T}}_i {\bf V}_{i-1} \; ,
\end{equation}
where the non-zero matrix elements of ${\mathcal{T}}_i$ are

\begin{eqnarray} \label{T}
t_{i1,1} & = & \langle \cos\theta\rangle_i \, ,
\; t_{i1,2} =  -\big\langle\frac{1}{\sin\theta} \big\rangle_{i-1}
\langle \sin\theta\cos\phi\rangle_i \, ,\nonumber \\
t_{i1,3} & = & \langle \sin\theta\cos\phi\rangle_i \, ,
\; t_{i1,4} = \langle \sin\theta\sin\phi\rangle_i \, ,\nonumber \\
t_{i2,1} & = & 1 \, , \nonumber \\
t_{i3,1} & = & \langle \cot\theta\cos\theta\rangle_i\, ,
\; t_{i3,2} =  -\big\langle\frac{1}{\sin\theta} \big\rangle_{i-1}  \langle
\cos\theta\cos\phi\rangle_i \, ,\nonumber \\
t_{i3,3} & = & \langle \cos\theta\cos\phi\rangle_i \, ,
\; t_{i3,4} = \langle \cos\theta\sin\phi\rangle_i \, ,\nonumber \\
t_{i4,2} & = & \big\langle\frac{1}{\sin\theta} \big\rangle_{i-1} \langle
\sin\phi\rangle_i \, ,
\nonumber \\
t_{i4,3} & = & -\langle \sin\phi\rangle_i \, ,
\; t_{i4,4} = \langle \cos\phi\rangle_i
\; .
\end{eqnarray}

Thus given the initial condition ${\bf V}_2$, which depends only on
$\rho_2(\theta)$, all successive ${\bf V}$'s can be calculated
recursively using eq.(\ref{rec}).
Let us discuss the case of a uniform stretch where
$\rho_i(\theta, \phi)$, and thence ${\mathcal{T}}_i$, does not depend
on $i$ (the sub-indices $i$'s will be omitted in this case). If the
left and right eigenvectors of ${\mathcal{T}}$, ${\bf W}_{\mu}$  and
${\bf W}^{\mu}$ respectively, corresponding to the eigenvalue
$\lambda_{\mu}$, form a complete basis set the general solution of
eq.(\ref{rec}) can be written as

\begin{equation}
{\bf V}_n = \sum_{\mu = 1}^4 \lambda_{\mu}^{n-2} {\bf W}_{\mu}\cdot
{\bf V}_2 \; {\bf W}^{\mu} \; .
\end{equation}
If all the eigenvalues are real and positive and if $\lambda \equiv
\max_{\mu=1,\dots,4} \{ \lambda_{\mu}\}$ then for large $n$
\begin{equation}
\langle \hat{t}_2 \cdot \hat{t}_{n+2}\rangle \sim \lambda^n \; ,
\end{equation}
and likewise for $\langle \hat{b}_2 \cdot \hat{b}_n\rangle $.  On
physical grounds, we expect that $\lambda < 1$, so that the correlation
functions decay exponentially with the distance measured along the
chain.  However it is quite common that some eigenvalues are complex.
Since the matrix $\mathcal{T}$ is real, complex eigenvalues occur in
pairs of complex conjugate values. If the pair $\lambda_{\pm}=\exp(\pm
i \chi - 1/\xi)$ ($\chi$ and $\xi$ are both real and positive)
corresponds to the maximum modulus eigenvalue, then at large $n$ we
get, for example, for the tangent-tangent correlation,

\begin{equation} \label{osci}
\langle \hat{t}_2 \cdot \hat{t}_{n+2}\rangle \sim
\cos(\chi_0 + n\chi)e^{-n/\xi} \; ,
\end{equation}
where $\chi_0$ depends on the initial conditions. Thus there is still
an exponential decay with a correlation length $\xi$ (in units of chain
bond length), but there is also an oscillatory modulation with another
length scale, $1/\chi$, which corresponds to short range order along
the chain (notice that, in one dimensional systems such as our chain,
long range order cannot occur if the interactions are short range
as in the present case). This type of behavior, with $1/\chi \sim 3.6$,
would be expected on a stretch of chain that adopts a
helical conformation with $3.6$ amino acids per turn.

We end this appendix with an example of such
behavior which can be worked out in full detail. For the case
$\rho_i(\theta,\phi) = \rho_i(\theta,-\phi)$ (which corresponds
to invariance under chirality flipping), and $\rho_i$ is
independent of $i$ for $i > 2$, then $t_{k,4} = t_{4,k} = 0 \; \textrm{with
}k \ne 4$. This implies that the matrix $\mathcal{T}$ becomes block
diagonal with an eigenvalue equal to $\langle\cos\phi\rangle$ and
$\langle {\bf x}\cdot\hat{b}_n\rangle$ decays exponentially with a
correlation length $-1/\ln\langle\cos\phi\rangle$. Furthermore, since
$t_{1,2}t_{3,3} = t_{1,3}t_{3,2}$, one eigenvalue is zero and the
remaining two are the solutions of the second order equation
$\lambda^2 + b \lambda + c = 0$ with
\begin{equation}
b = -t_{1,1} - t_{3,3} \;, \quad
c = t_{1,1}t_{3,3}-t_{1,3}t_{3,1}-t_{1,2} \; .
\end{equation}

Thus, if $b^2 - 4c > 0$, the two solutions are real and the
tangent-tangent correlation decays exponentially to zero. On the other
hand, if $b^2 - 4c < 0$, the two solutions are complex conjugate
of each other, as
described above, and in the particular case we are considering, i.e.
$\rho_i(\theta,\phi) = \rho_i(\theta,-\phi)$, one finds for all
$n$, that
\begin{equation}
\langle \hat{t}_1 \cdot \hat{t}_{n+2}\rangle =
\frac{\cos(\chi_0 + n\chi)}{\cos\chi_0}e^{-n/\xi} \; ,
\end{equation}
where $\xi=-2/\ln c$, $\chi=\arccos\left(-b/2\sqrt{c}\right)$, and
$\chi_0$ depends on the initial conditions.

\eject

\begin{thebibliography}{99}
\bibitem{Secondary} The importance of local information is
underscored by the fact that many schemes for
secondary structure prediction do moderately well with just local
sequence information.  See also the discussion presented by
Srinivasan and Rose\cite{Linus} on the importance of local
interactions in protein folding.
\bibitem{note13} An interesting problem for thick polymers in the
continuum limit is the measure associated with the curve
corresponding to the viable centerline of a thick tube. Indeed,
for the Edwards' model, the first term in the effective energy in
Eq. (\ref{EM}) is used to define a measure on the ensemble of all
continuous curves embedded in three-dimensional space, the Wiener
measure.  It is well known that this gives a non-zero weight to
all curves which are continuous but not differentiable ($C^0$
curves in the standard notation). In particular, the curves for
which the tangent can be defined are a subset of zero measure in
this ensemble. It can be proved rigorously (see e.g. Ref.
\cite{gonz}) that if a curve is a viable centerline for a tube of
non zero thickness, then it must satisfy more stringent smoothness
requirements, namely it should be $C^{1,1}$, i.e. it is
differentiable and its derivative is a Lipschitz function (see
\cite{gonz} and Refs. therein).  An interesting problem, which we
do not consider here, is the definition of a suitable measure in
the continuum limit for such thick polymer centerlines.
\end{thebibliography}
\end{document}